\documentclass[]{aastex631}
\newcommand\rosat{{\it ROSAT\/}}

\newcommand\xmm{{\it XMM-Newton}}
\newcommand\XMM{{\it XMM-Newton}}

\newcommand\nustar{{\it NuSTAR}}
\newcommand\nice{{\it NICER}}

\newcommand{\fluxcgs}{ergs~s$^{-1}$~cm$^{-2}$}  
\newcommand{\lumcgs}{ergs~s$^{-1}$}

\newcommand{\mycomment}[1]{}

\newcommand{\src}{CJ2056} 
\def\amin{\ifmmode^{\prime}\else$^{\prime}$\fi}
\def\asec{\ifmmode^{\prime\prime}\else$^{\prime\prime}$\fi}
\def\simgt{\lower.5ex\hbox{$\; \buildrel > \over \sim \;$}}
\def\simlt{\lower.5ex\hbox{$\; \buildrel < \over \sim \;$}}

\usepackage{amsmath}
\usepackage{comment}
\usepackage{soul}

\begin{document}

\title{A Broadband X-ray Investigation of Fast-Spinning Intermediate Polar CTCV J2056--3014}  

\author[0000-0001-8586-7233]{Ciro Salcedo}
\affiliation{Columbia Astrophysics Laboratory, Columbia University, New York, NY 10027, USA}

\author[0000-0002-9709-5389]{Kaya Mori}
\affiliation{Columbia Astrophysics Laboratory, Columbia University, New York, NY 10027, USA}

\author[0000-0002-6653-4975]{Gabriel Bridges}
\affiliation{Columbia Astrophysics Laboratory, Columbia University, New York, NY 10027, USA}

\author[0000-0002-3681-145X]{Charles J. Hailey}
\affiliation{Columbia Astrophysics Laboratory, Columbia University, New York, NY 10027, USA}

\author[0000-0002-7004-9956]{David A. H. Buckley}
\affiliation{South African Astronomical Observatory, P.O Box 9, Observatory, 7935 Cape Town, South Africa}
\affiliation{Department of Astronomy, University of Cape Town, Private Bag X3, Rondebosch 7701, South Africa}

\author[0000-0002-6211-7226]{Raimundo Lopes de Oliveira}
\affiliation{Departamento de F\'isica, Universidade Federal de Sergipe, Av. Marechal Rondon, S/N, 49100-000, S\~ao Crist\'ov\~ao, SE, Brazil}
\affiliation{Observat\'orio Nacional, Rua Gal. Jos\'e Cristino 77, 20921-400, Rio~de~Janeiro, RJ, Brazil}

\author[0000-0001-8722-9710]{Gavin Ramsay}
\affiliation{Armagh Observatory and Planetarium, College Hill, Armagh, BT61 9DG, UK}

\author[0000-0003-0397-4655]{Anke van Dyk}
\affiliation{Department of Astronomy, University of Cape Town, Private Bag X3, Rondebosch 7701, South Africa}

\begin{abstract}
We report on \XMM, \nustar, and \nice\ X-ray observations of CTCV J2056-3014, a cataclysmic variable (CV) with one of the fastest-spinning white dwarfs (WDs) at $P=29.6$ s. While previously classified as an intermediate polar (IP), \src\ also exhibits the properties of WZ-Sge-type CVs, such as dwarf novae and superoutbursts. With \xmm\ and \nice, we detected the spin period up to $\sim 2$ keV with $7\sigma$ significance. We constrained its derivative to $|\dot{P}| < 1.8 \times 10^{-12}$ s/s after correcting for binary orbital motion. The pulsed profile is characterized by a single broad peak with $\sim25$\% modulation. \nustar\ detected a four-fold increase in unabsorbed X-ray flux coincident with an optical flare in November 2022. The \XMM\ and \nice\ X-ray spectra in 0.3–10 keV are best characterized by an absorbed optically-thin three-temperature thermal plasma model ($kT = 0.3, 1.0$ and 4.9 keV), while the \nustar\ spectra in 3-30 keV are best fit by a single-temperature thermal plasma model ($kT = 8.4$ keV), both with Fe abundance $Z_{\rm Fe}/Z_\odot = 0.3$. \src\ exhibits similarities to other fast-spinning CVs, such as low plasma temperatures, and no significant X-ray absorption at low energies. As the WD's magnetic field strength is unknown, we applied both non-magnetic and magnetic CV spectral models ({\tt MKCFLOW} and {\tt MCVSPEC}) to determine the WD mass. The derived WD mass range ($M = 0.7\rm{-}1.0\, M_\odot$) is above the centrifugal break-up mass limit of $0.56 M_\odot$ and consistent with the mean WD mass of local CVs ($M \approx 0.8\rm{-}0.9 M_\odot$).
\end{abstract} 

\section{Introduction} \label{sec:intro}
Cataclysmic variables (CVs) are white dwarf (WD) binary systems where mass accretion via Roche-lobe overflow from a late-type main sequence companion leads to X-ray emission. CVs are the most common interacting compact binaries and are potential progenitors for type 1a supernovae, making their study vital for testing theories of stellar evolution. With the advent of more sensitive optical and X-ray surveys (e.g., ZTF, {\it eROSITA}), a rare class of fast-spinning CVs (FSCVs) with $P_{\rm spin}$ $<$ 50 seconds has been revealed, which are distinct from regular accretion-powered CVs in various ways. 

Among the few FSCVs discovered so far, AE Aqr,  LAMOST J024048+195226 (J0240), WZ Sge, and V1460 Her stand out for their exotic properties (Table \ref{tab:src_types}) \citep{1938AN....265..345Z, Pelisoli2022, 1957ApJ...126...23G, 2017NewA...52....8K}. AE Aqr, whose WD is spinning at 33.1 s, was identified as the first propeller CV, whose spinning WD magnetosphere ejects incoming gas particles, as evidenced by its rapid spin-down rate, highly variable H-$\alpha$ lines (which indicate outflow winds) and lack of accretion disk signatures in its Doppler tomograms \citep{Wynn1997}. Its WD mass was measured at M = $0.63 \pm 0.05\; M_{\odot}$, representing the first FSCV mass measurement \citep{aqrmass}. J0240 is another propeller CV system recently discovered  with the fastest-spinning WD detected so far \citep[$P_{\rm spin} = 24.9$ s; ][] {Pelisoli2022, PretoriusPropeller}. WZ~Sge-type CVs exhibit occasional outbursts in the optical band, indicating more variable mass accretion, including dwarf novae outbursts and the much rarer superoutbursts. The most notable source in this category is WZ Sge, with two short periods at 27.87 and 28.96 s, in the optical and X-ray bands, respectively \citep{Nucita2014}. Due to the anomalous difference between the optical and X-ray periods, these periods have not been securely associated with WD rotation \citep{Nucita2014}. Another FSCV, V1460 has a spin period of 39 s and demonstrates typical IP properties but has not been studied in the X-ray band \citep{PelisoliIP}.

Although each of the FSCVs demonstrates distinguishing features within their class, they share some common properties. First, they are usually classified as intermediate polars (IPs) due to their asynchronous spin and orbital periods. IPs are a subclass of magnetic CVs, defined by non-synchronized orbits, with WD magnetic fields ($B\sim0.1-10$~MG) strong enough to truncate the inner accretion disk. While most of the known IPs are bright and copious emitters of hard X-rays \citep[$L_{\rm X} \simgt 10^{33}$ erg\,s$^{-1}$ and $kT \sim 20{\rm-}40$ keV; ][]{Mukai2017}, the FSCVs we mention are significantly fainter in the X-ray band ($L_{\rm X} \sim   10^{30}\rm{-}10^{31}$~\lumcgs). Thus they have also been classified as low luminosity IPs (LLIPs). Lastly, AE Aqr and WZ Sge are presently spinning down with $\dot{P} \sim10^{-13} \rm{-} 10^{-11} $ [s\,s$^{-1}$], in stark contrast to regular IPs, most of which are slowly spinning up near the spin equilibrium \citep{Patterson2020}. V1460 Her's spin derivative has not been measured, but its upper limit of $|\dot{P}| < 3 \times 10^{-14}$ s/s is smaller than that of the other known FSCVs \citep{PelisoliIP}. 

FSCVs provide a unique opportunity to study the WD interior structure and the evolution of WD spins and magnetic fields. For example, the relationship between WD mass and spin is a fundamental question for exploring the interior structure of WDs. Theoretical studies of WD stability predict that WDs can spin as fast as $\sim1$-s and stay intact if its mass approaches the Chandrasekhar limit of $M \sim 1.4 M_\odot$ (e.g., \citet{Boshkayev2013}). Fast-spinning WDs must be sufficiently massive to avoid being broken up by centrifugal force without having enough mass to destabilize the core, inducing inverse $\beta$-decay and pyronuclear reactions \citep{Otoniel2020}.  
In contrast to the FSCVs, isolated WDs are relatively slower rotators with spin periods ranging from hours to years  \citep{Ferrario2015}, although some spin as fast as $P \sim 70 $ s as a result of WD mergers \citep{Kilic2021}. 
While mass accretion from companion stars in CVs represents the primary avenue of spinning up WDs, it is unknown how their WD spins and magnetic fields have evolved due to their limited population. 
While several theoretical studies have attempted to understand their emission mechanisms and evolutionary paths in a more unified context \citep{Lyutikov2020}, significant progress waits on the discovery of more FSCVs.

\begin{table*}[h] 
\small 
\caption{A selected list of FSCVs with $P_{\rm spin} \lesssim 50$~s}
\centering 
\begin{tabular}{lccccc}
\hline\hline
Source name & $P$ [sec] & $\dot{P}$ [s/s] & $P_{\rm orb}$ [hr]  & $L_{\rm X}$ [erg\,s$^{-1}$] & Comments \\ 
\hline 
AE Aqr & 33.1$^a$ &  $6\times10^{-14}$ $^a$ & 9.88$^b$ & $4\times10^{30}$ $^c$ & Propeller \\ 
J0240$^{1}$  & 24.9$^d$ &  --- & 7.34$^e$ & --- & Propeller   \\ 
WZ Sge & 27.87 \& 28.96$^2$ & $8\times10^{-12}$ $^f$ & 1.35$^g$ & $2\times10^{30}$ $^c$& Superoutbursts  \\
V1460 Her  & 39$^h$ &  $<|3\times10^{-14}|$ $^h$ & 4.99$^i$ & $1\times10^{30}$ $^j$ & IP \\ 
CTCV J2056 & 29.6$^k$ &  $ < |2 \times 10^{-12}|$ & 1.76$^l$&  $2\times10^{31}$ &  --- \\ 
\hline\hline 
\end{tabular} 

1) J0240 has not been observed in the X-ray band\\
2) It is still debated if the twin periods are associated with WD rotation \citep{Nucita2014}]\\
a) \cite{Li2016}; b) \cite{Dejager1994}; c) \cite{2020A&A...641A.136W}; d) \citep{Pelisoli2022}; e) \cite{2014ApJS..213....9D}; f) \cite{Patterson1980}; g) \cite{1962PASP...74...66K}; h) \cite{PelisoliIP}; i) \cite{Ashley2020}; j) \cite{Evans2020}; k) \cite{Oliveira}; l) \cite{Augusteijn2010}

\label{tab:src_types}
\end{table*}

CTCV J2056$-$3014 (\src\ hereafter) is another addition to the family of FSCVs. \src\ is a nearby IP discovered by the Cal{\'a}n-Tololo optical survey, with an X-ray counterpart detected by \rosat\ \citep{Augusteijn2010}. Remarkably, follow-up optical and X-ray observations detected a spin period of 29.6 sec, currently making it the 2nd fastest spinning WDs ever detected \citep{Oliveira}, only after the propeller CV J0240. Moreover, the orbital period of 1.76 hrs is below the period gap ($P_{\rm orb}\sim$2--3 hrs), where the binary orbit decays via GW radiation rather than magnetic braking. 
Other than WZ Sge, this is the only known FSCV below the period gap, implying that \src\ may be in a much later stage of CV evolution than the other FSCVs \citep{Zharikov2015, Rodrigues2023}. 
The X-ray timing and spectral features of \src\ have been previously studied using \xmm\ data \citep{Oliveira}. An initial 18--ks \xmm\ observation identified a 29.6--s period, which was confirmed with a reanalysis of optical observations of \src. Like other FSCVs, \src\ was found to be faint in the X-ray band with $L_{X,0.3-12 \mathrm{ keV}} = 1.8 \times 10^{31}$ \lumcgs. 
In the optical band, high-speed photometry observations at the South African Astronomical Observatory revealed occasional signal period fluctuation or splitting around the 29.6-sec spin period (van Dyk in preparation). 
The cause of the puzzling optical spin variability, similar to those observed from WZ Sge \citep{Nucita2014}, is unknown but will be discussed in our companion paper by van Dyk et al. 
Furthermore, a handful of dwarf novae events (DNe) and a superoutburst followed by a reflaring event have been detected in the optical band \citep{Hameury2022}. While DNe are frequent optical outbursts caused by disk instability, superoutbursts are brighter and longer, originating from thermal-tidal instability in the disk caused by resonances \citep{Hameury2020}. 
By conventional definitions, \src\ could be classified as a WZ Sge-type CV, a subset of the SU UMA-type CVs, given the DNe, superoutburst, and superhump observed in the optical band, making it the 3rd IP in this category \citep{Byckling2010, Hameury2022}. Its cataclysmic properties suggest that \src\ is a distinct FSCV powered by highly variable mass accretion. While these optical characteristics suggest \src\ may be a non-magnetic CV (nmCV), where the accretion disk reaches the WD surface, the X-ray properties observed so far indicated that it is an IP with tall accretion columns \citep{Oliveira}. 

We present further X-ray and optical observations in this paper and companion paper (van Dyk et al.), respectively. To fully characterize the broadband X-ray properties of \src, we performed new \xmm, \nustar\, and \nice\ observations. This paper presents a more extensive X-ray timing and spectral analysis of \src. \S\ref{sec:data} describes all X-ray observations of \src\ and data reduction methods. In \S\ref{sec:time}, we search for the spin period in all X-ray data sets, investigate its stability over time, and constrain the spin evolution. Folded X-ray light curves are explored in different energy bands. In \S\ref{sec:spec}, we characterize the phase-averaged and phase-resolved X-ray spectra with phenomenological spectral models. The plasma temperatures, Fe abundance, and atomic lines are well constrained by fitting the broadband X-ray data. In \S\ref{sec:mass}, we determine the WD mass range with the more sophisticated X-ray spectral models, assuming that \src\ is either a non-magnetic or magnetic CV. In \S\ref{sec:disc}, we discuss the implications of our X-ray analysis, including the WD stability condition of \src, compared with other FSCVs. In \S\ref{sec:conc}, we conclude the paper with the future of exploring the rare class of FSCVs.

\section{X-ray Observations and Data Reduction} 
\label{sec:data}

\begin{figure}[!t]
\begin{center}
\includegraphics[width=18cm]{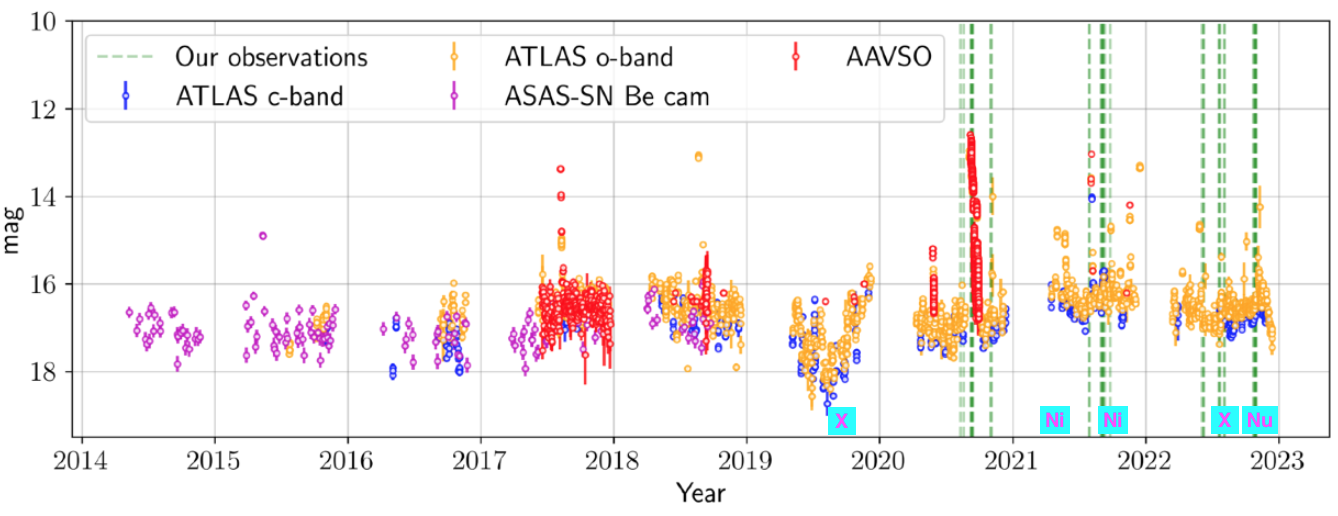}
\caption{Long-term optical lightcurve of \src, overlaid with the SALT observations indicated by vertical, dashed green lines. The \xmm, \nice\, and \nustar\ observations are denoted by "X," "Ni," and "Nu" cyan boxes on the bottom. }
\label{fig:optical_lc} 
\end{center} 
\end{figure}

\src\ was observed in the X-ray band five times between 2019 and 2022 (Table \ref{tab:obstimeline}). Figure \ref{fig:optical_lc} shows the optical light curve of \src\ over the last decade, overlaid with all X-ray observation dates. Besides the routine optical monitoring provided by ASSASN \citep{ASS1, ASS2}, ATLAS \citep{atl}, and AAVSO \cite{aavso}, dedicated optical observations at SAAO were conducted and reported in our companion paper (van Dyk et al.).

\src\ was observed by \xmm\ twice on October 24, 2019 \citep[18 ks;][]{Oliveira} and on October 23, 2022 (54 ks). 
The source extraction region was a circle of $r=20''$ around the source position, while an annular region of $r=30\rm{-}50''$ was used for background extraction. All \xmm\ data sets were reduced using SAS 19.1, using \textit{emchain} and \textit{epchain} for GTI filtering for MOS and pn modules, respectively, with \textit{evselect} for region extraction \citep{2004ASPC..314..759G}. The 2019 observation yielded 0.3--10 keV count rates of 0.103, 0.096, and 0.429 cts/s  for the MOS1, MOS2, and pn modules, respectively. Meanwhile, the 2022 observation yielded count rates of 0.108 (MOS1), 0.110 (MOS2) and 0.528 (pn) cts/s. We collected $\sim13,000$ and $\sim40,000$ net counts for MOS and pn data in 0.3--10 keV after combining the 2019 and 2021 \xmm\ observation data. 

\src\ was observed by \nice\ for 53 ks exposure on July 25, 2021. \nice\ observations were again performed on November 6, 2021, with 20 ks of total exposure. Both data sets were reduced using \textbf{nicerl2}, which generates cleaned and reduced event and spectrum files. Background model spectra were produced using \textbf{nibackgen3C50} \citep{3c50}. 
The \nice\ observations yielded $\sim67,000$ and $\sim25,000$ total counts in the 0.5 - 1.5 keV band, resulting in count rates of 1.27 and 1.22 cts/s, respectively, beyond which background events dominated. 

A 21 ks \nustar\ observation was conducted on November 4, 2022. We processed the \nustar\ data using {\tt nupipeline} \citep{Harrison2013}. We extracted source events from a $r=30''$ circular region around the target position. An annular region at $r=100\rm{-}180''$ was used for extracting background events. The \nustar\ data, with FPMA and FPMB module events combined, yielded $\sim$ 3,700 net counts in the 3--30 keV band after background subtraction, with respective count rates of 0.088 and 0.082 cts/s. We found that the \nustar\ data were heavily contaminated by background photons above 30 keV.

\begin{deluxetable}{lccc}[!h]
\tablecaption{X-Ray observation log of \src}
\tablehead{ \colhead{Date}   
&
\colhead{ObsId}  
& 
\colhead{Telescope} 
&  
\colhead{Exposure (ks)} }
\startdata
2019-10-24  & 0842570101 & \xmm\ $^1$  & 18    \\
2021-07-25/29  & 459201020* & \nice\ & 53   \\
2021-11-05/06  & 459201030* & \nice\ & 20   \\
2022-10-23  & 0902500101 & \xmm\ $^1$  & 54    \\
2022-11-04  & 30801006002 & \nustar\ & 21 \\
\enddata
$^1$ This observation was carried out in Prime-Full Window Mode \\
$^*$ NICER observations are collected in several successive observations sharing the same obsID prefix\\
\label{tab:obstimeline}
\end{deluxetable}

\section{Timing Analysis} \label{sec:time}
    
All X-ray observation data were analyzed with the {\tt Stingray} software package \citep{matteo_bachetti_2022_6394742} and {\tt Hendrics} \citep{hendrics}
for our timing study. First, we applied barycentric correction to all extracted source events. The SAS command \textbf{barycen} was used for the 2019 and 2022 \XMM\ data. MOS 1 and 2 data were combined for timing analysis. Meanwhile, the \nice\ and \nustar\ source events were corrected with the \textbf{barycorr} command from HEASOFT Version 6.25 \citep{heasoft}. We applied energy filtering using {\tt Hendrics} command-line tools.     

\subsection{Periodicity search} 

We began our analysis with the $H$-test for $n = 1,2,3$ harmonic components using the weighted $Z^2_n$ function through  \textbf{z\_n\_search} in {\tt Stingray}. Our periodicity search swept over a frequency range of $f = 10^{-3} \rm{-} 10^{2}$ Hz using the reciprocal of the observation length as the frequency step size for each dataset, yielding the most significant detections ($>\; 7\sigma$ significance) of the 29.6 sec spin period at $n$=3 in the \xmm\ and \nice\ observations. The significance and false-alarm probabilities of the peak signals presented account for the number of trials which is given by the length of the frequency range divided by the frequency step-size of each search. In the \xmm\ periodograms, sub-harmonic signals are present. Using fewer harmonic components, i.e., $n=1$ and  $n= 1, 2$ resulted in detecting the spin period at similar significance (e.g., for the first \nice\ observation, $Z^2_1 = 989.29$ and $Z^2_2 = 1019.96$), without additional sub-harmonic signals. Meanwhile, we detected the spin signal in the \nice\ observations in the same energy band (0.3--2 keV). Due to aperiodic variability, we did not detect its sub-harmonic with \nice\, regardless of how many harmonics were summed.

\begin{figure}[ht]
\includegraphics[width=.333\linewidth]{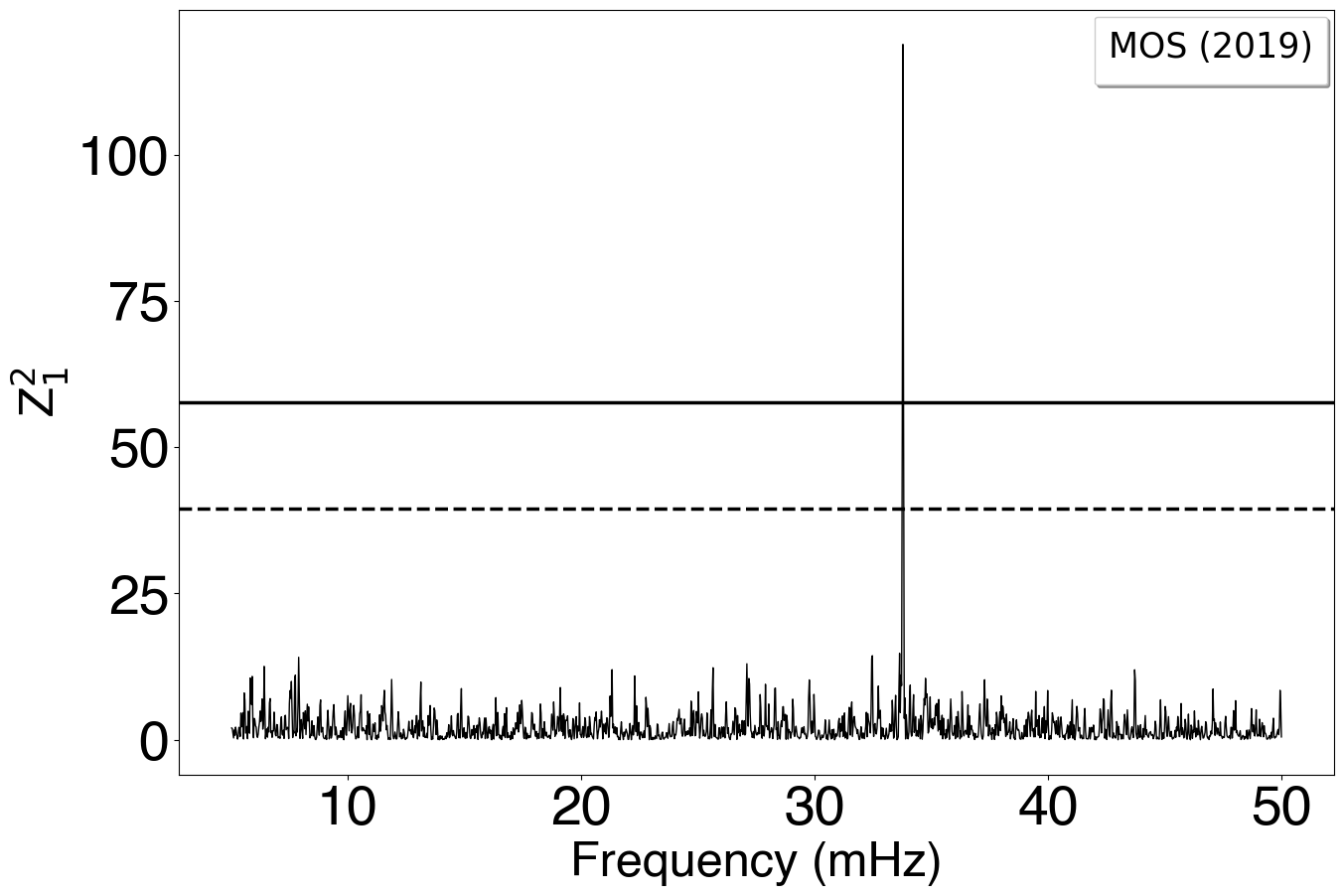}\hspace{0cm}
\includegraphics[width=.34\linewidth]{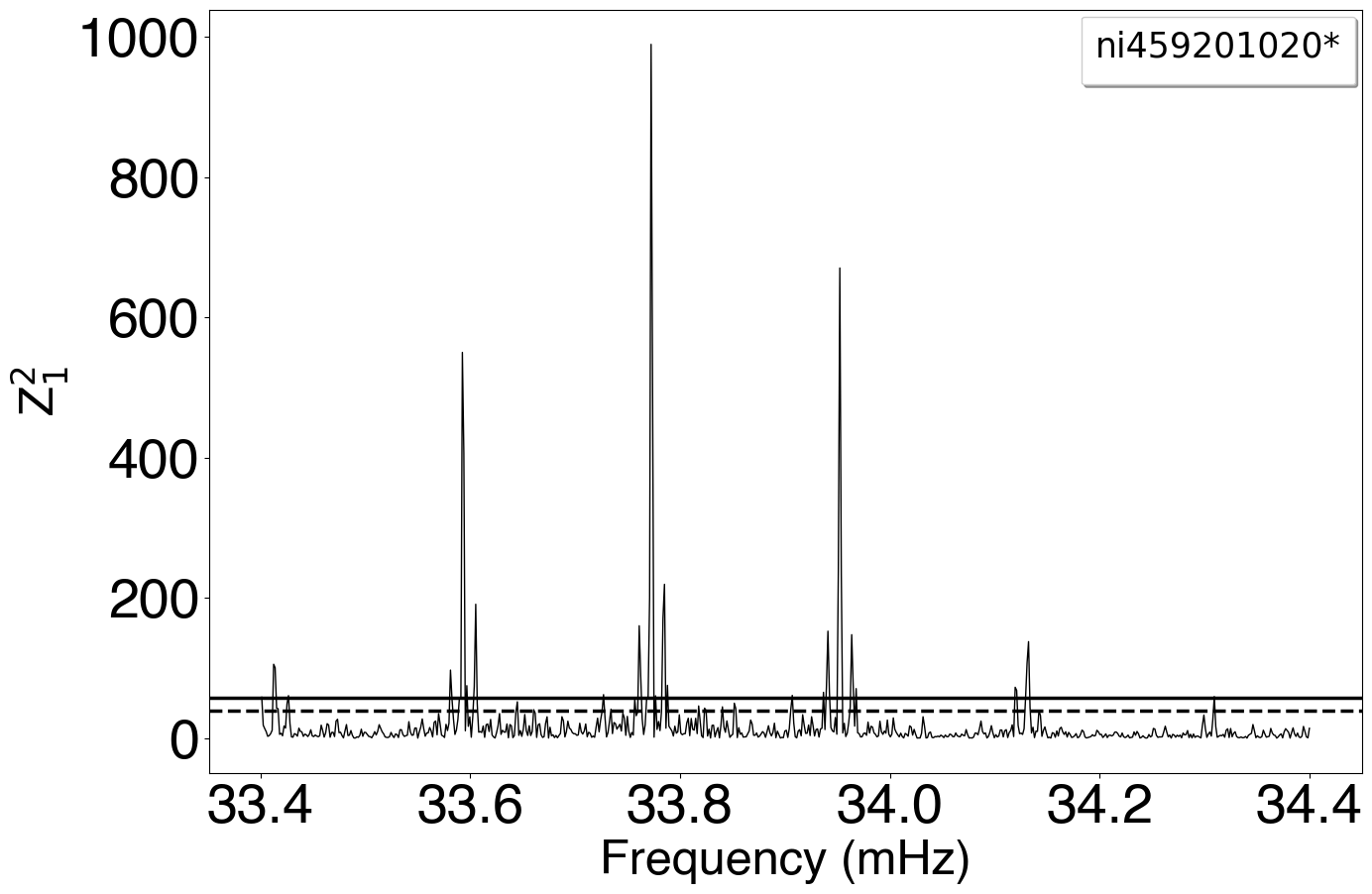}
\includegraphics[width=.333\linewidth]{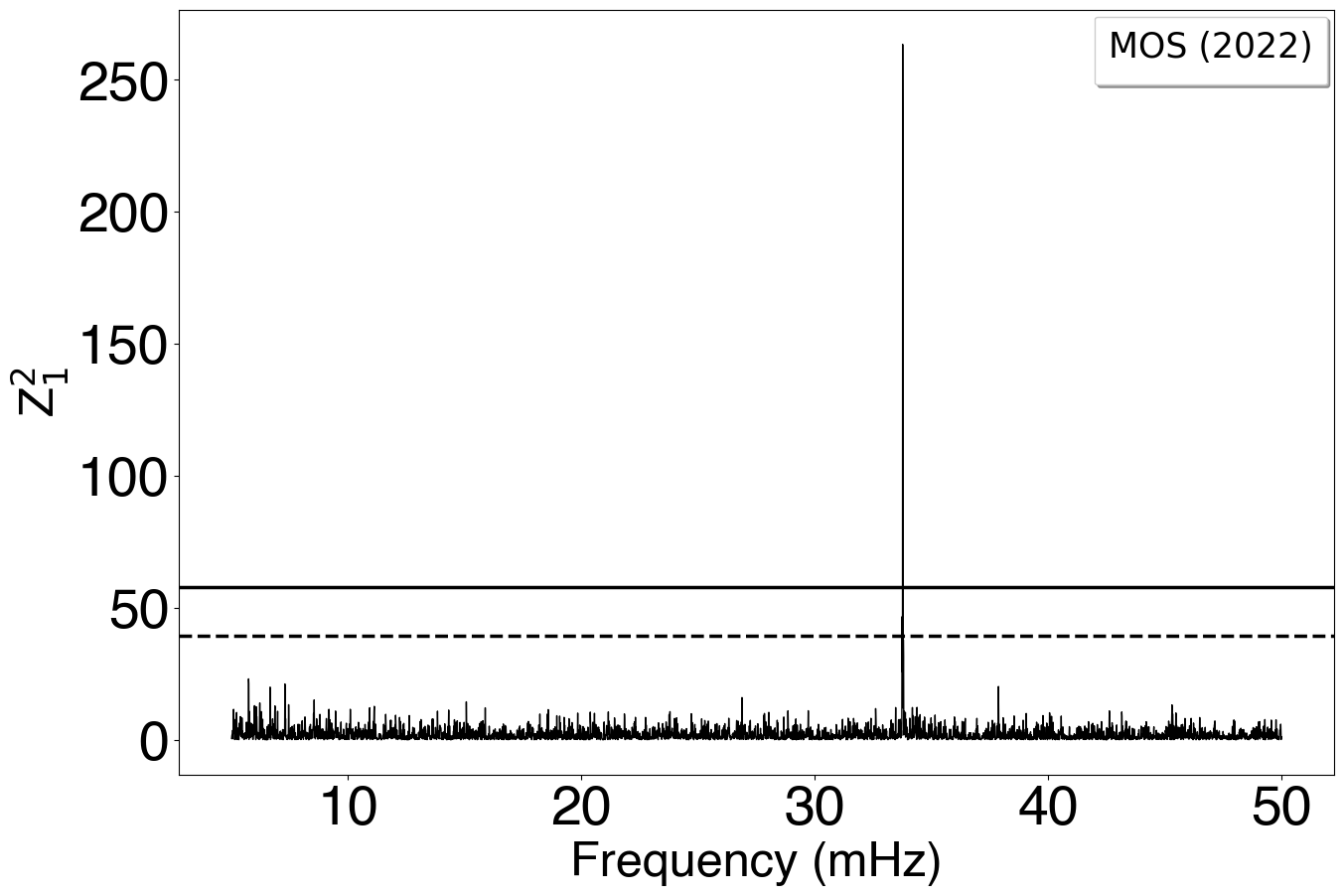}\\
\includegraphics[width=.333\linewidth]{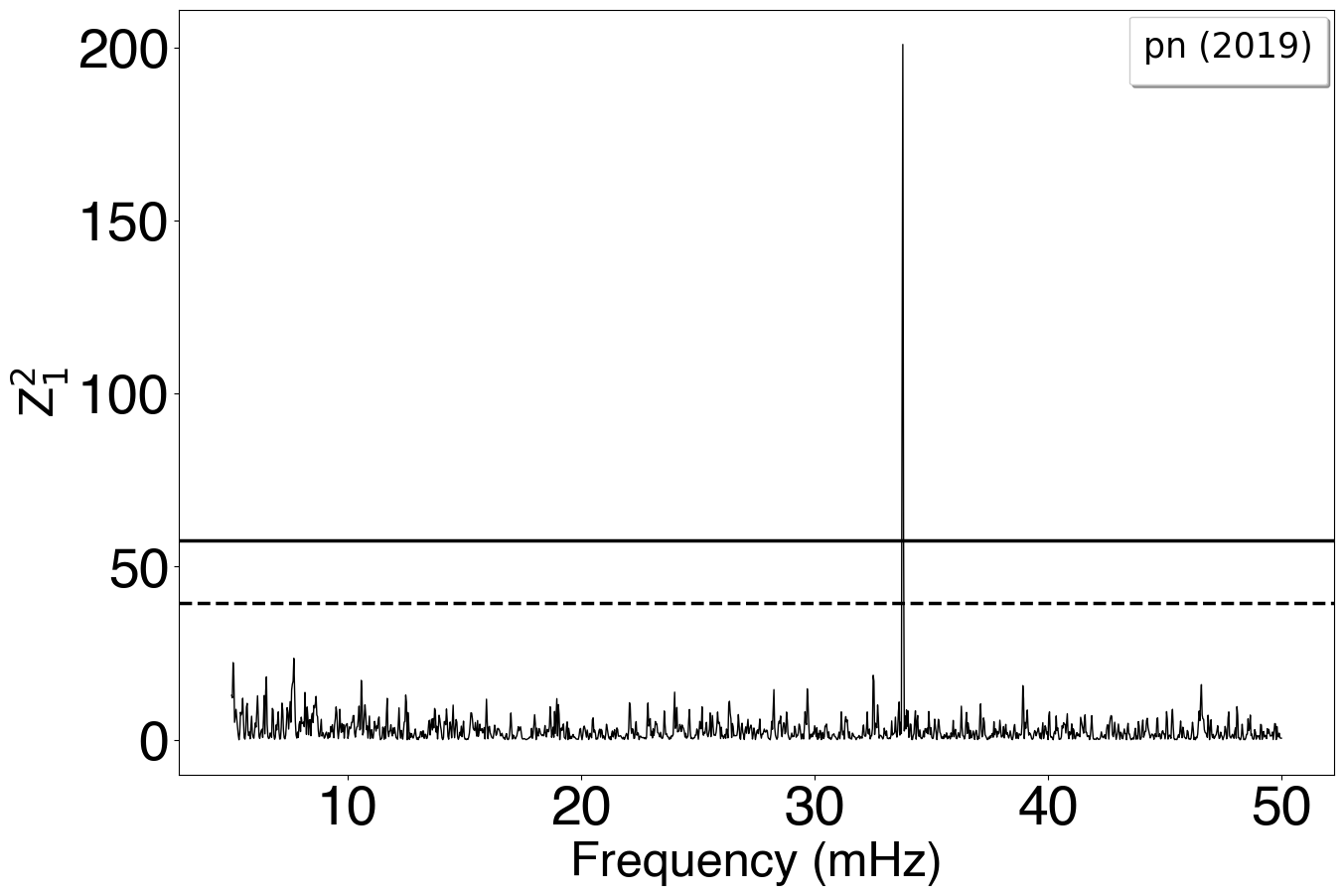}\hspace{0cm}
\includegraphics[width=.333\linewidth]{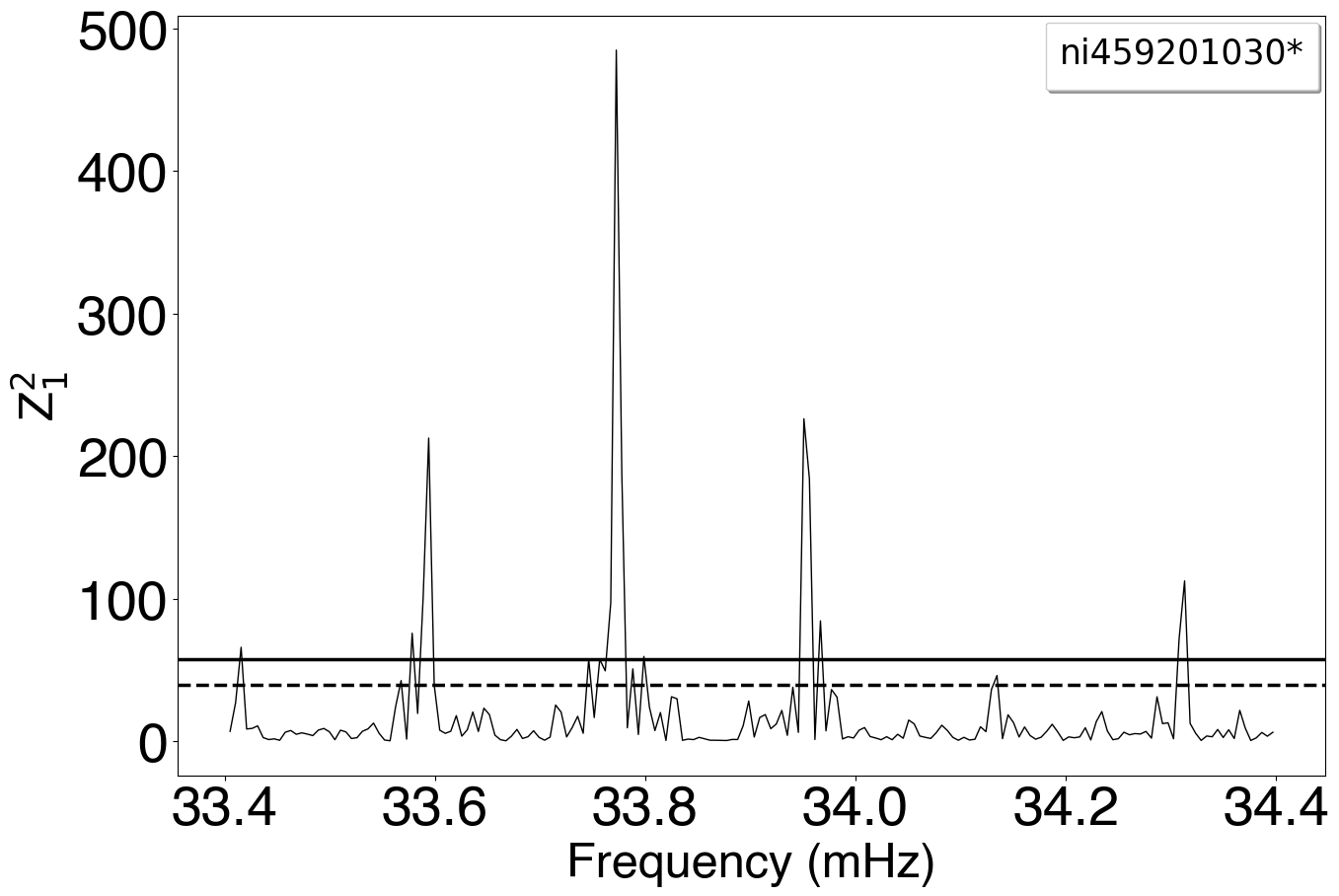}
\includegraphics[width=.333\linewidth]{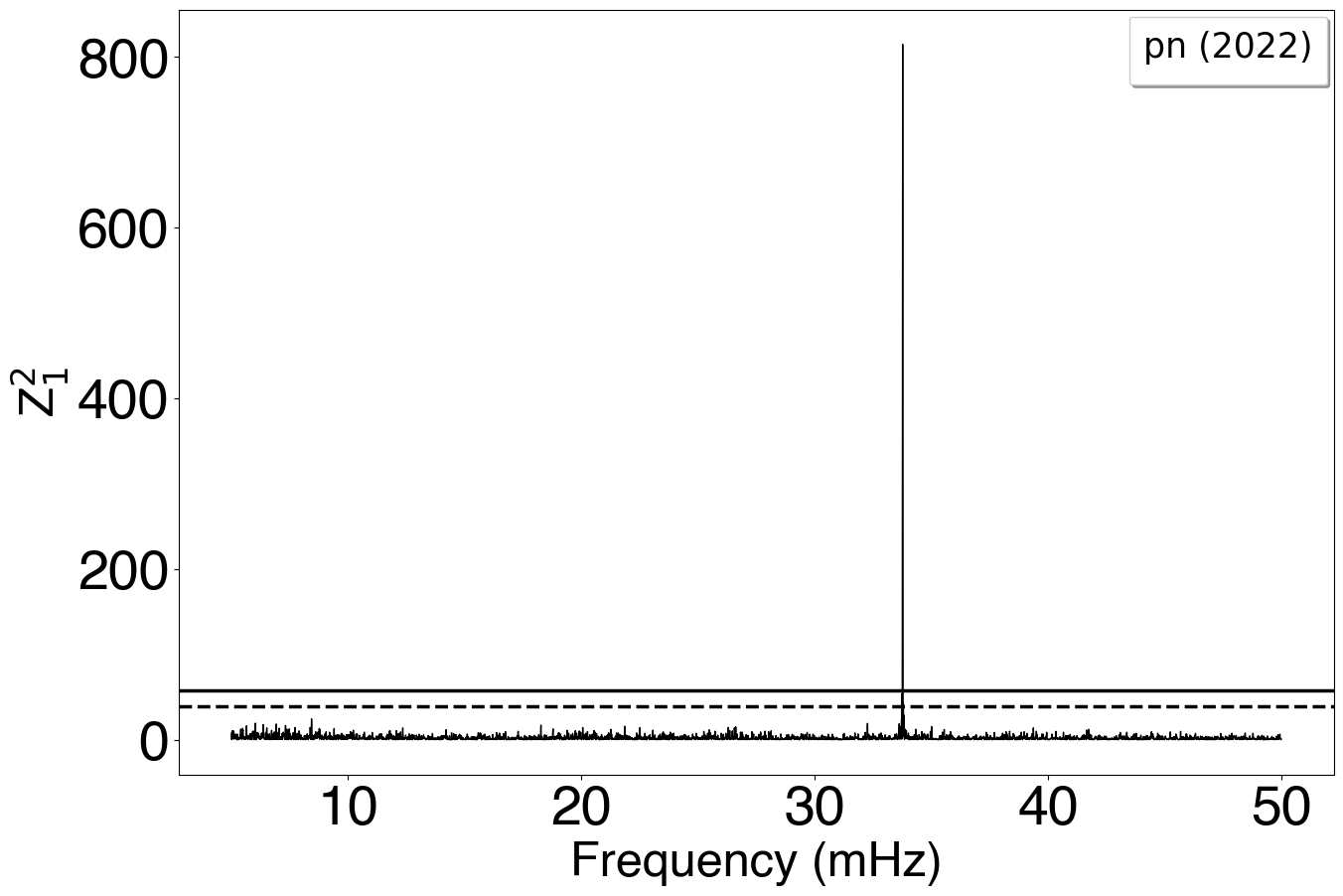}
  \caption{$Z^2_1$ soft X-ray (0.3 - 2 keV) periodograms. The top row from left to right: 0842570101 MOS, 459201020* \nice, and 0902500101 MOS. The bottom row from left to right: 0842570101 pn, 459201030* \nice, and 0902500101 pn. In each periodogram, 3-$\sigma$ and 5-$\sigma$ significance levels are shown by solid and dashed lines, respectively. The most significant peak is in the region of P $\sim$ 29.6 seconds. While the 5 mHz to 25 mHz range is excluded from the \nice\ periodograms due to substantial aperiodic variability, for the XMM observations, the harmonic of the 29.6 sec signal is highly significant at 59.2 seconds. }
\label{fig:z2tests} 
\end{figure}

At higher energy bands above 2 keV, the spin period was more weakly detected in the \xmm\ and \nice\ data; the significance of the peak signal of the $Z^2_1$ searches for each dataset never exceeded $4-\sigma$. As the $Z^2$ statistic of the peak frequency was only negligibly increased by summing out to three harmonics, we used the $Z^2_1$ test in a narrower frequency band ($f = 5\rm{-}50$ mHz) around the spin period to characterize the peak and its uncertainty without the spurious subharmonics (Figure: \ref{fig:z2tests}). In the narrow-band $Z^2$ data, we fit the 29.6-sec peak with a Gaussian function and found its width of $\Delta f < 0.007$ mHz. The frequency resolution is consistent with the Nyquist limit corresponding to a reciprocal of the observation length. Later, we constrained the spin period more accurately based on the two \nice\ observations as \nice\ provides better temporal resolution ($< \mu$s) than \xmm.

\begin{figure}[ht] 
\begin{center} 
\includegraphics[width=8.5cm]{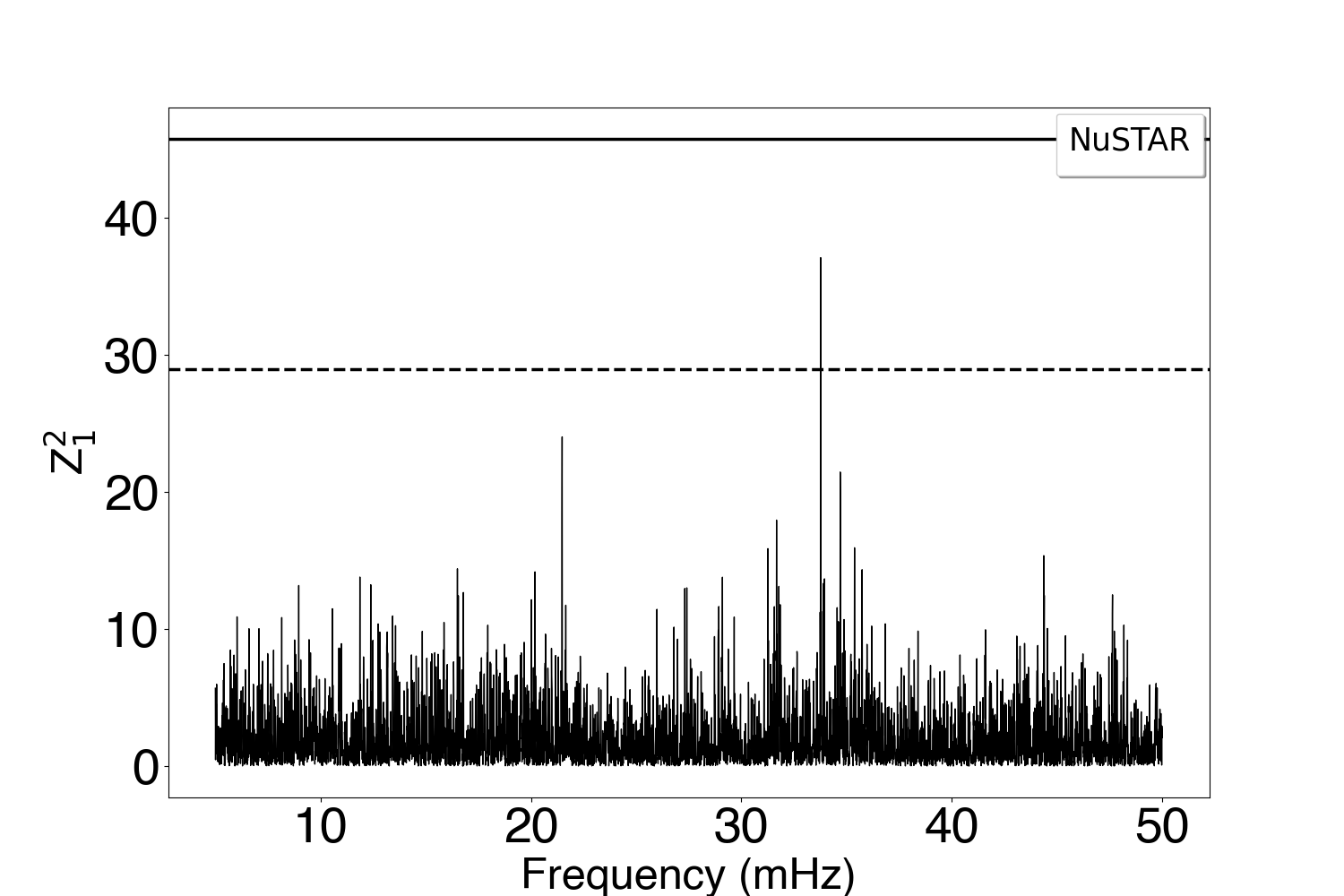}
\includegraphics[width=8.5cm]{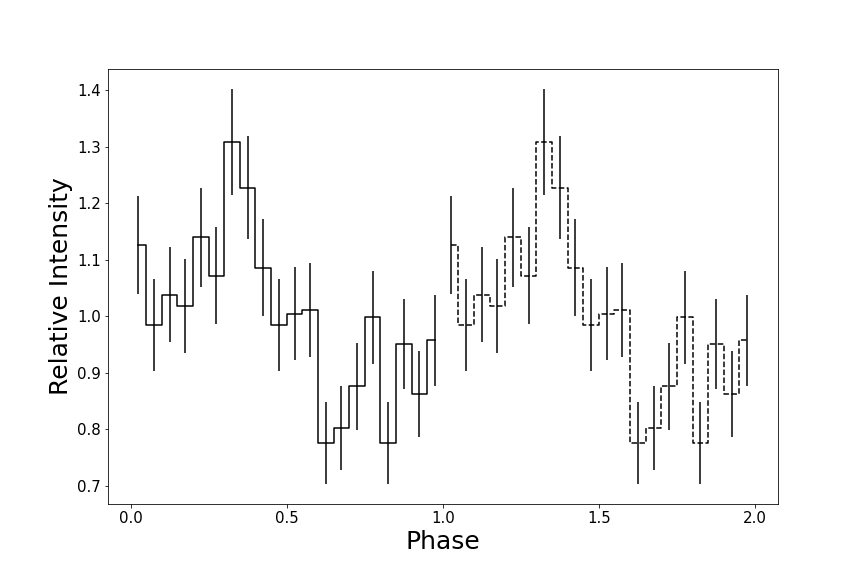}

\end{center} 
\caption{\small 
 Left:30801006002 FPMA+FPMB joint $Z^2_1$ search (5 mHz - 50 mHz). The 3-$\sigma$ and 5-$\sigma$ significance levels are shown by solid and dashed lines, respectively.
 Right: The pulse profile of the folded \nustar\ events at 29.6 sec. Relative intensity is defined by the mean count rate of the folded events. 1-$\sigma$ error bars are included.} 
\label{fig:nustar_time} 
\end{figure}

We performed the same timing analysis above 3 keV using the \nustar\ data. As we performed the $Z^2_1$ test in the same frequency band ($f = 5 \rm{-} 50$ mHz) as the \xmm\ timing analysis, we found that a peak at the spin period was detected at the 4.0$-\sigma$ level. (Figure \ref{fig:nustar_time}). 

Table \ref{tab:timing} summarizes all of our spin period measurements. While our results are consistent with the 2019 \xmm\ observation findings \citep{Oliveira}, joint \nice\ data analysis from the two observations separated by $\sim$ 3 months reduced the spin period uncertainty significantly compared to the \xmm\ results.

\subsection{Folded X-ray lightcurves and hardness ratio profiles}

\begin{figure}[ht]
\includegraphics[width=.333\linewidth]{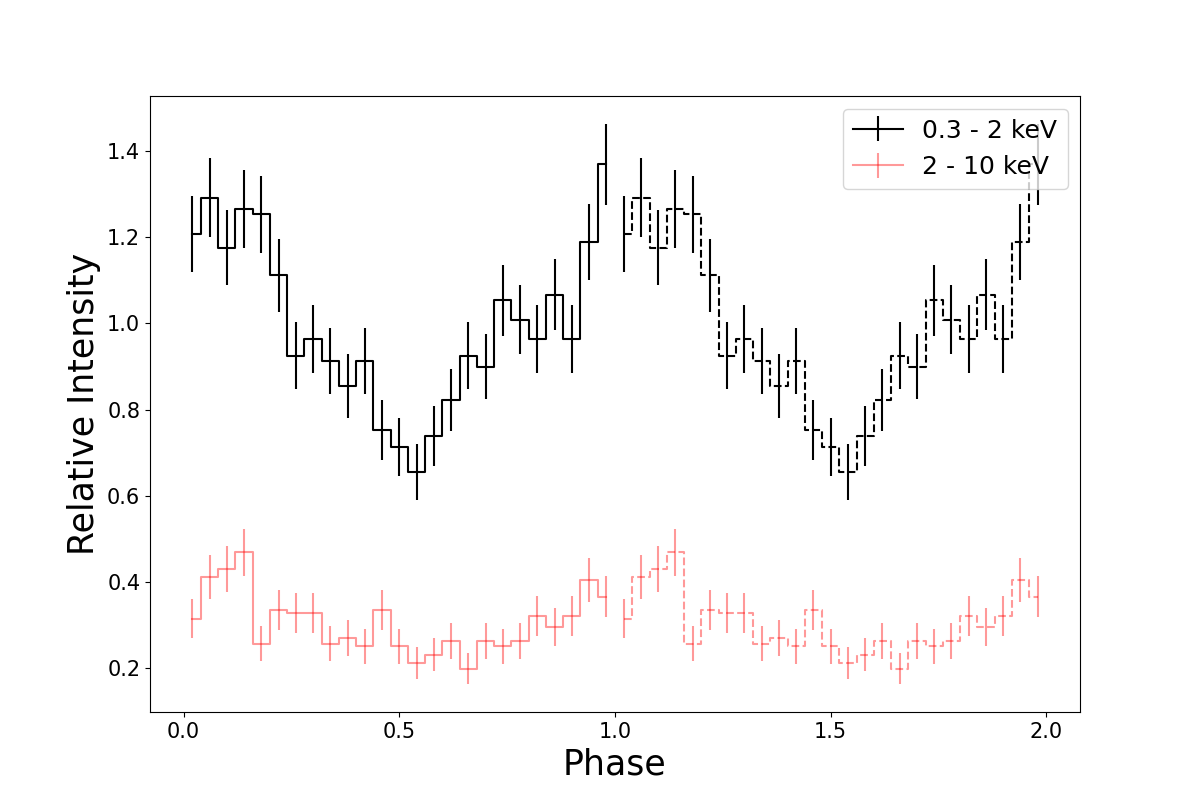}
\includegraphics[width=.333\linewidth]{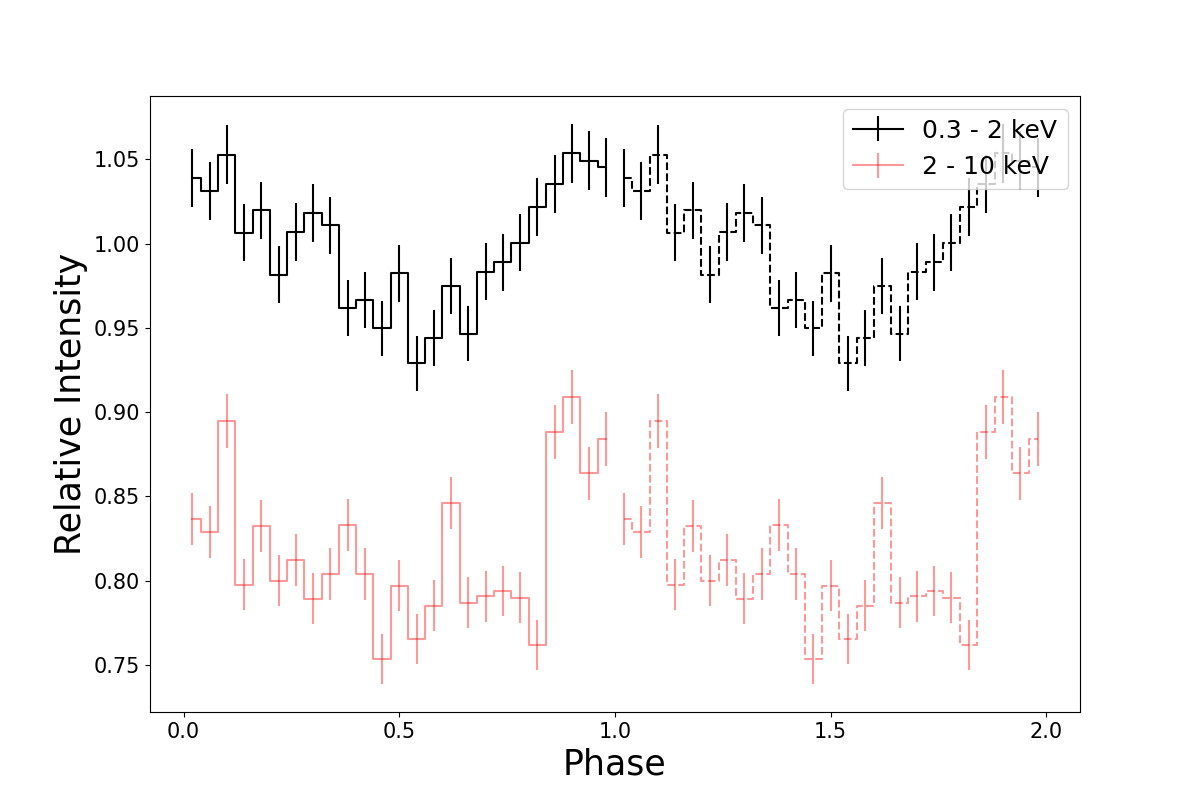}
\includegraphics[width=.333\linewidth]{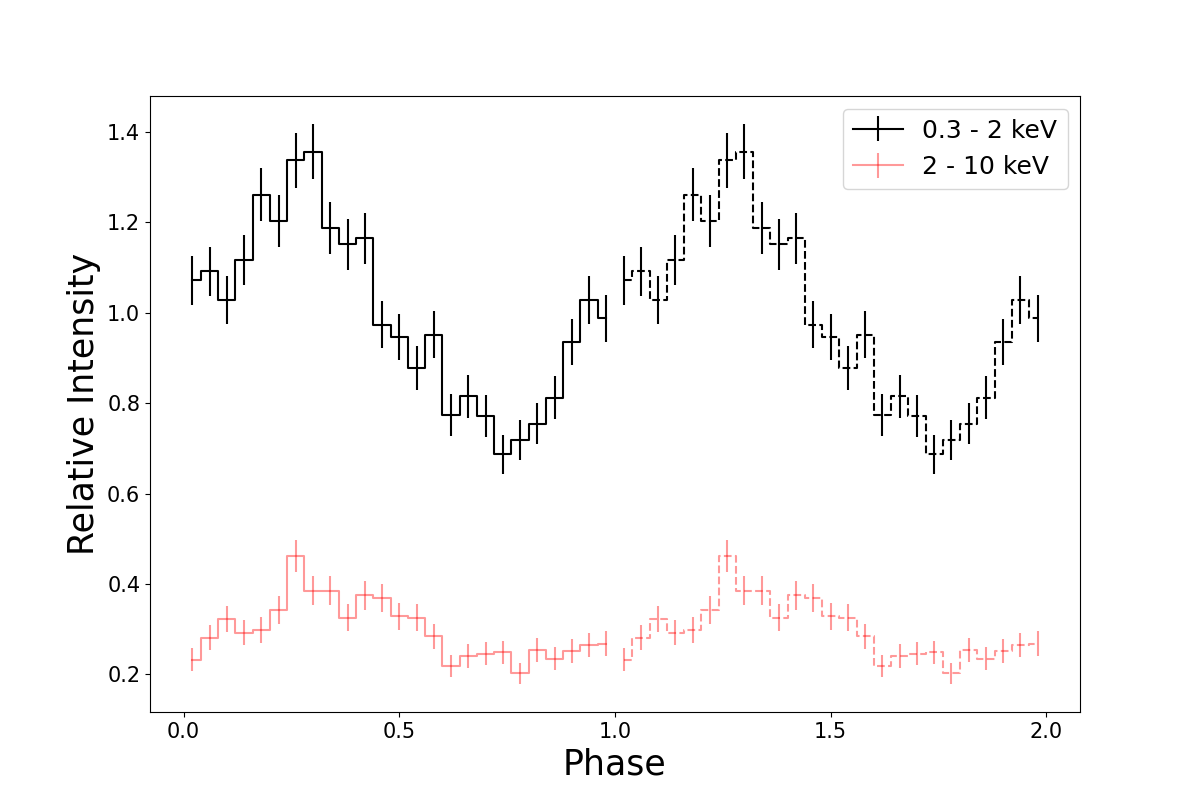}\\
\includegraphics[width=.333\linewidth]{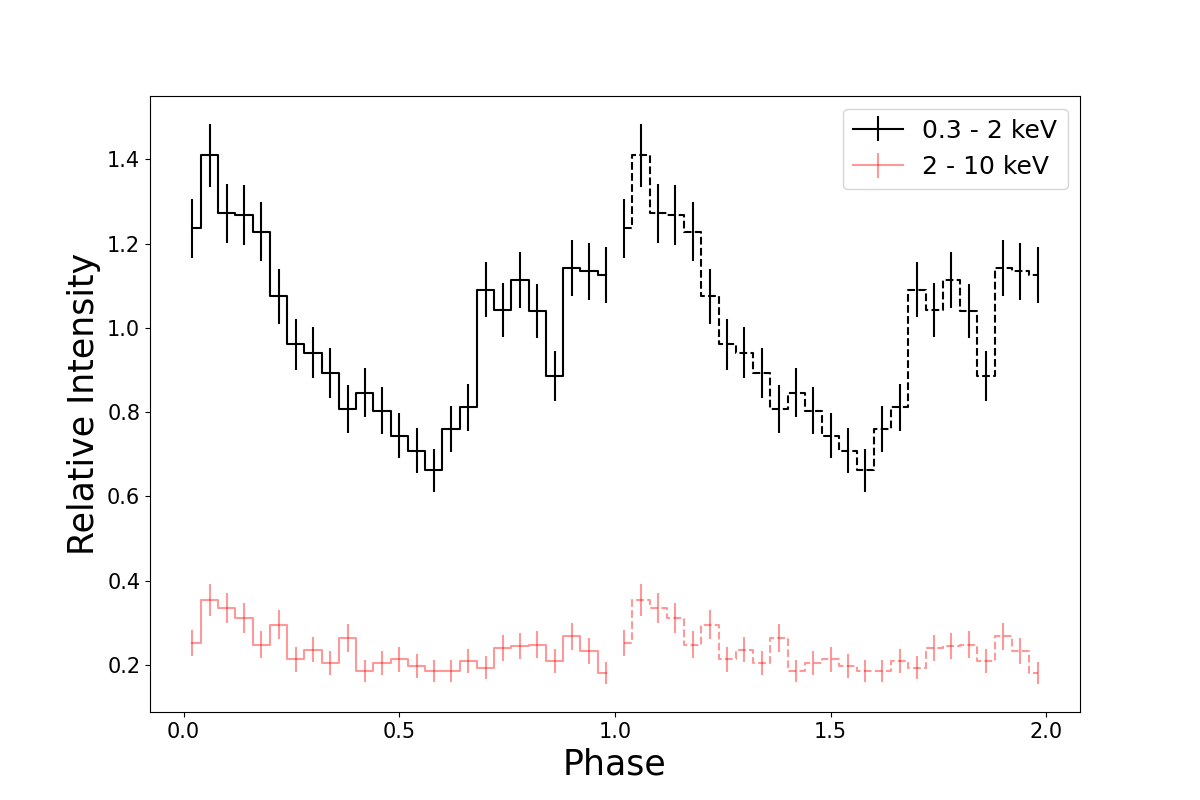}
\includegraphics[width=.333\linewidth]{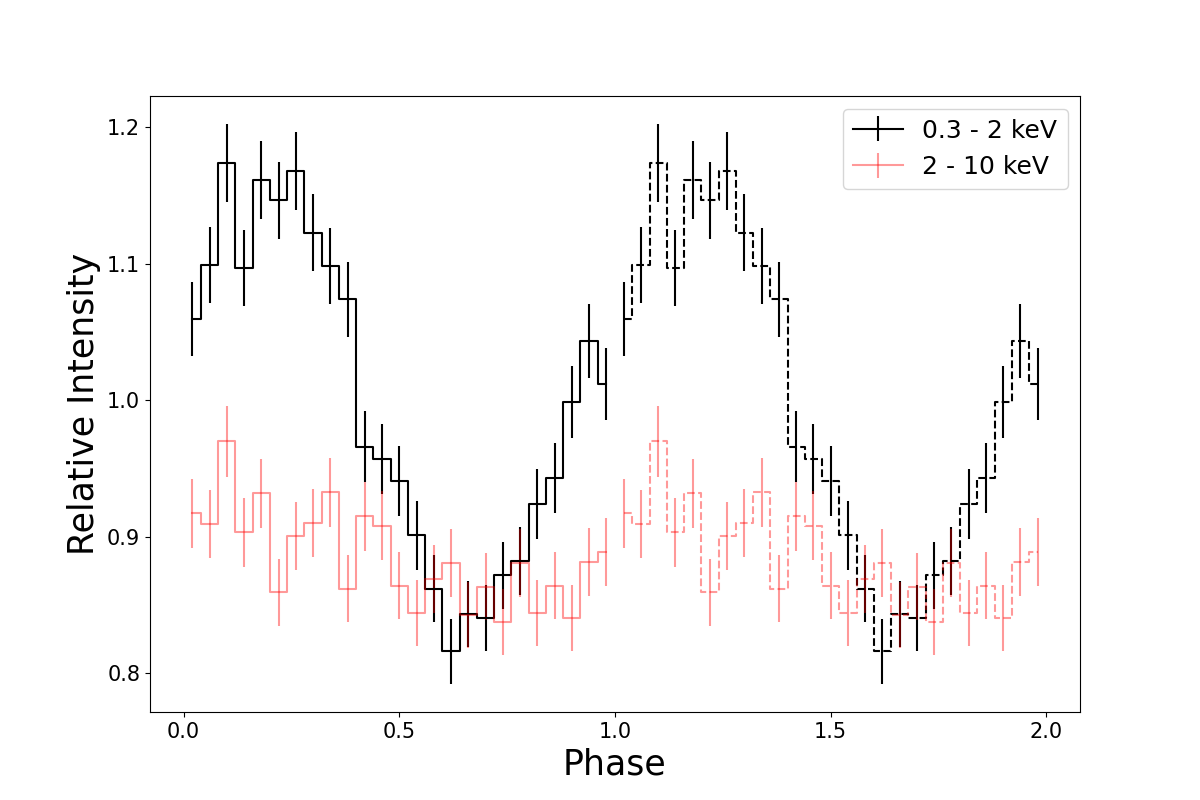}
\includegraphics[width=.333\linewidth]{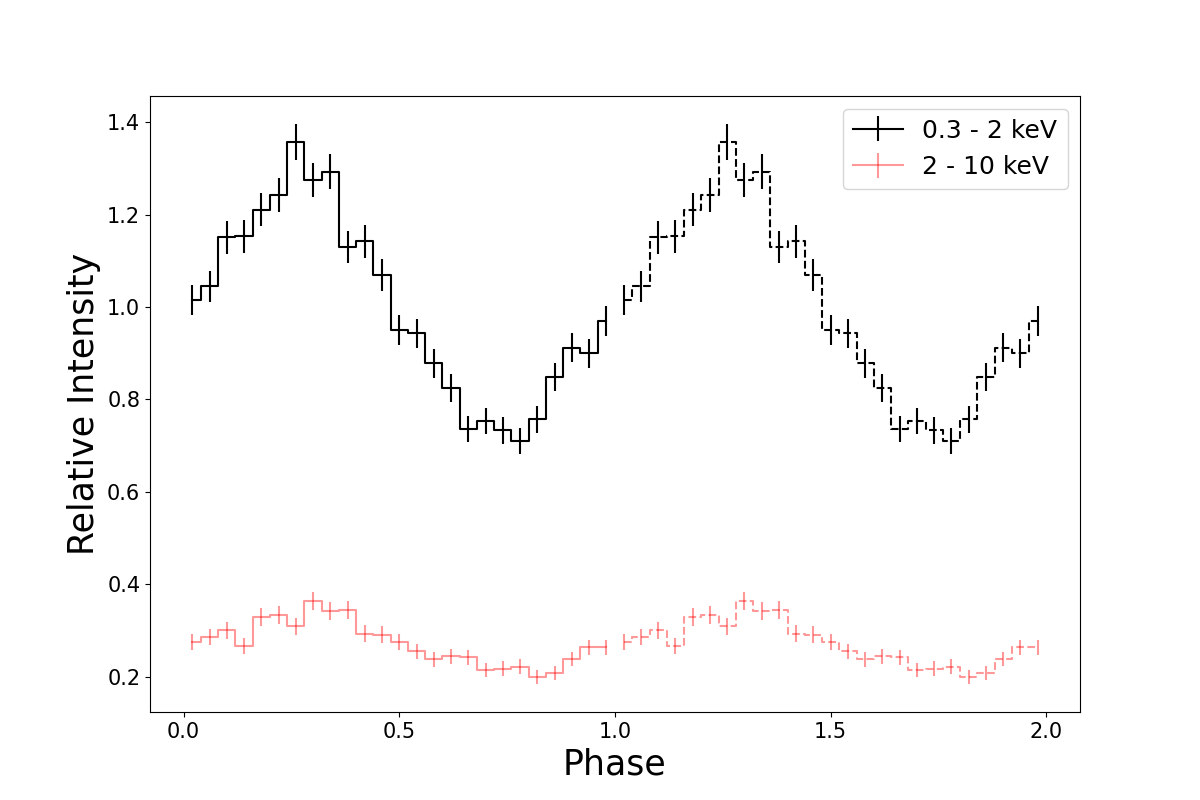}
  \caption{Pulse profiles of folded source events in the soft (0.3 - 2 keV) and hard (2 - 10 keV) X-ray band. he top row from left to right: 0842570101 MOS, 459201020* \nice, and 0902500101 MOS. The bottom row from left to right: 0842570101 pn, 459201030* \nice, and 0902500101 pn. Relative intensity is defined as the fraction of the mean count rate of the soft X-ray profile. 1-$\sigma$ error bars are included.}
  \label{fig:pulse}
\end{figure}

Following the spin period detection, we produced pulsed profiles folded at 29.6 sec for each X-ray observation. Although no significant periodicity was detected above 2 keV, we generated folded lightcurves in the 0.3--2 keV and 2--10 keV bands, using {\tt Stingray}'s \textbf{fold\_events} with 20 bins per cycle. The pulse fraction, as defined by {\tt stingray} is $\frac{a-b}{\mathrm{x}}$, where $a$ and $b$ are the minimum and maximum of the profile and $x$ is the mean counts across the phase bins \citep{matteo_bachetti_2022_6394742}. The soft X-ray profiles exhibit a single broad peak with a pulse fraction between 15 \% and 25 \% (Table \ref{tab:timing}). The spin modulation was more prominent in the \XMM\ observations than the  \nice\ observations, which were subjected to higher background contamination and thus a higher mean counts compared to the amplitude of modulation (Figure \ref{fig:pulse}). With the \nustar\ data, we generated a 3--10 keV pulsed profile folded over the 29.6 sec period and found modulation with a pulse fraction of 16.2 \%. We also produced X-ray lightcurves folded by $P=59.2$ sec, which exhibit two symmetric non-overlapping pulses. It establishes that the 29.6 s signal represents the fundamental spin period, confirming the results of \cite{Oliveira}.

\begin{figure}[ht] 
\begin{center} 
\includegraphics[width=8.5cm]{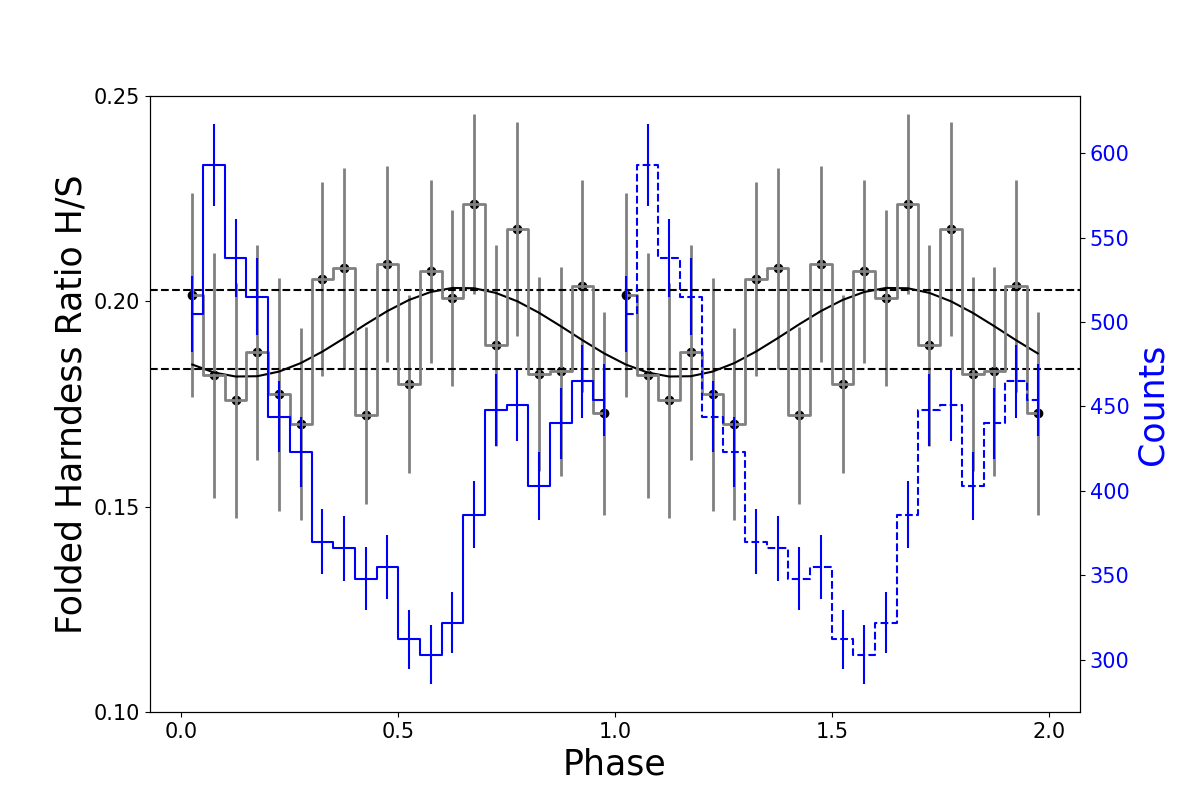}
\includegraphics[width=8.5cm]{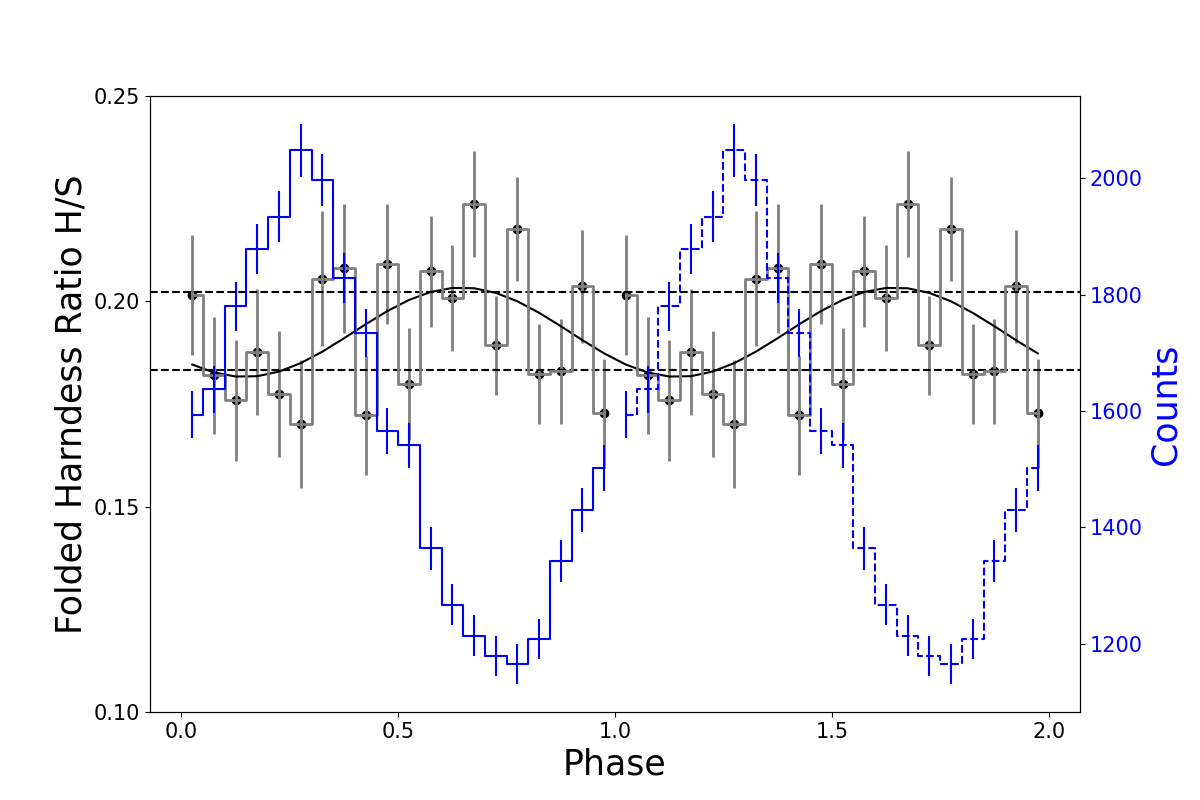}

\end{center} 
\caption{\small 
The hardness ratio folded at 29.6 sec with 20 bins per cycle for the 2019 pn (left) and 2022 pn (right) observations. The dotted lines represent the 3-$\sigma$ significance line for deviation from the mean. Overlaid are the best-fit sinusoidal curves in black. In blue, we present the folded light curve of each observation in the 0.3 - 5 keV band to demonstrate the anticorrelation between the hardness ratio and the luminosity.} 
\label{fig:hard} 
\end{figure}

Using the \xmm\ pn data from both \XMM\ observations, we generated hardness ratio curves folded over the 29.6-sec spin period with 20 phase bins. We compared the source counts between the 2--5 keV and 0.3--2 keV. The hardness ratio curves for the 2019 and 2022 \xmm\ observations demonstrated only weak modulations (Figure \ref{fig:hard}) possibly correlated with the X-ray flux modulation -- the X-ray emission became softer when the source was brighter. For better visualization, we overlaid the best-fit sinusoidal functions in the hardness ratio curves and the 3-$\sigma$ significance lines deviating from the mean hardness ratios. 
The sinusoidal function fit to the 2019 data had an amplitude of $N = 0.013 \pm 0.006$ with $\chi^2_\nu$ = 0.64 for 18 degrees of freedom (d.o.f.). Then, the spin modulation of the hardness ratio is not significant. The fit to the 2022 \xmm\ data, with nearly four times more source counts, produced $\chi^2_\nu$ = 0.95 for 18 d.o.f. with amplitude $N = 0.009 \pm 0.004$. This observation was less contaminated by background events, and the spin signal appeared with more than 3 times the $Z^2_1$ power in the new \xmm\ observation. In this extended observation, we detected hardness ratio modulation at 3-$\sigma$.

\subsection{The spin period stability and derivative}

\begin{figure}[ht] 
\begin{center} 
\includegraphics[width=8.7cm]{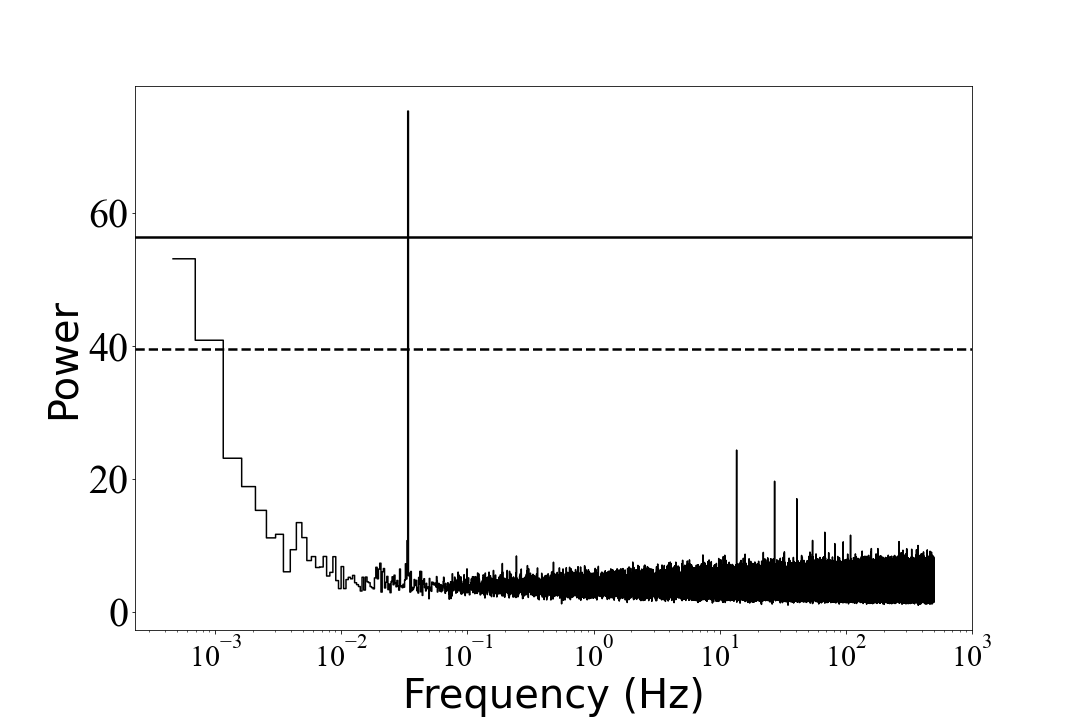}
\includegraphics[width=8.7cm]{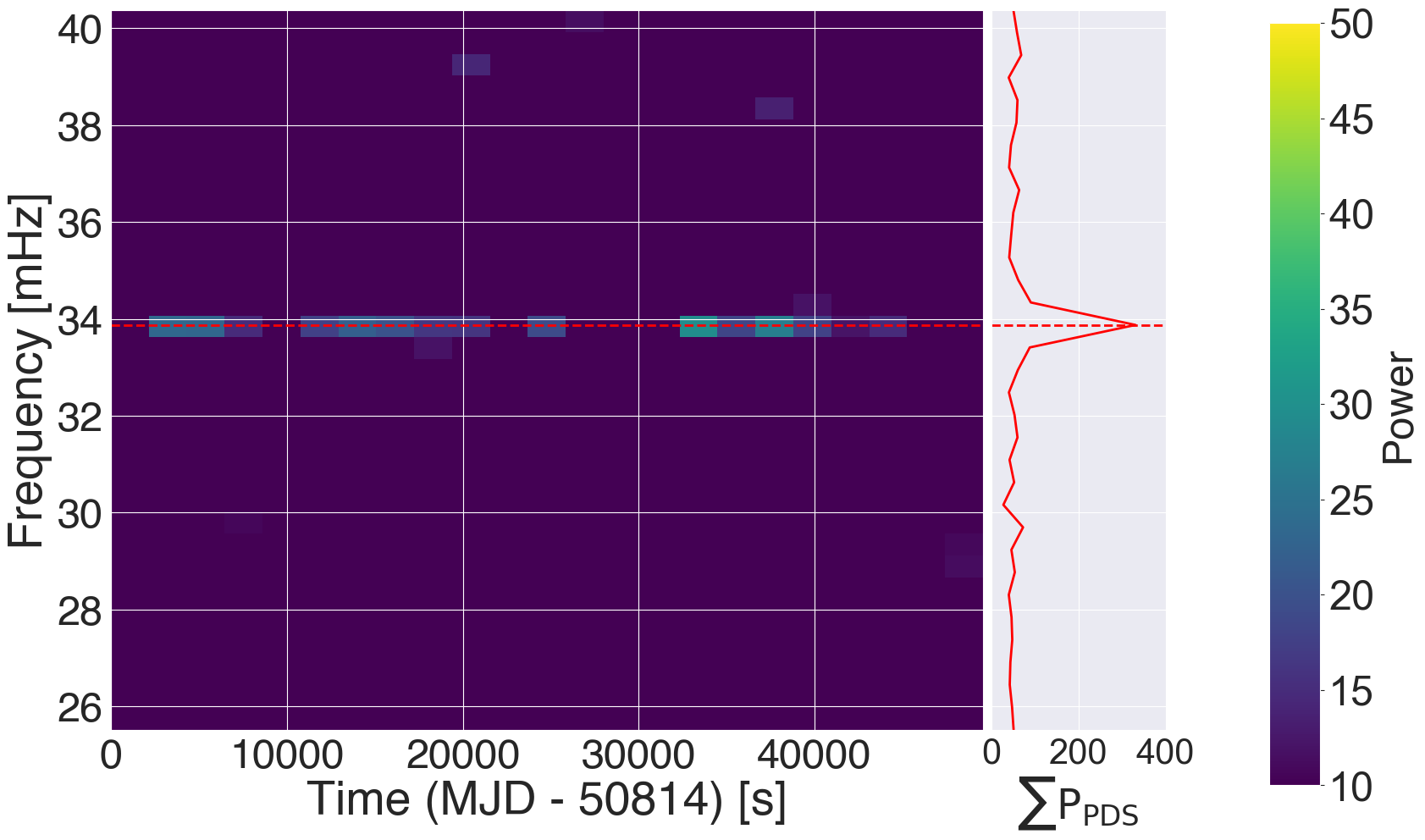}

\end{center} 
\caption{\small 
 Left: A PDS periodogram obtained from the 2022 \xmm-pn data in the 0.3 - 2 keV band, demonstrating a strong peak at $f \sim 33.7 \, \mathrm{mHz}$. The 5-$\sigma$ and 3-$\sigma$ significance lines are shown by the solid and dashed lines, respectively. Right: 2022 \XMM\ -EPIC pn dynamical power spectrum periodogram. The measured peak from the PDS is shown in green.} 
\label{fig:Best_PDS} 
\end{figure}

To investigate the stability of the spin period, we created dynamical power spectra in 0.3--2 keV from the \xmm\ pn observation data in 2022, as it provides the longest net exposure time and the longest continuous intervals of observation in this energy band among all X-ray observations (Figure \ref{fig:Best_PDS} right panel). We found that the 29.6-sec spin period was stable during the 54 ks X-ray observation in contrast to the occasional variability of the WD spin observed in the optical band (van Dyk et al. in prep). 

\begin{figure}
    \centering
    \includegraphics[width=8.5cm]{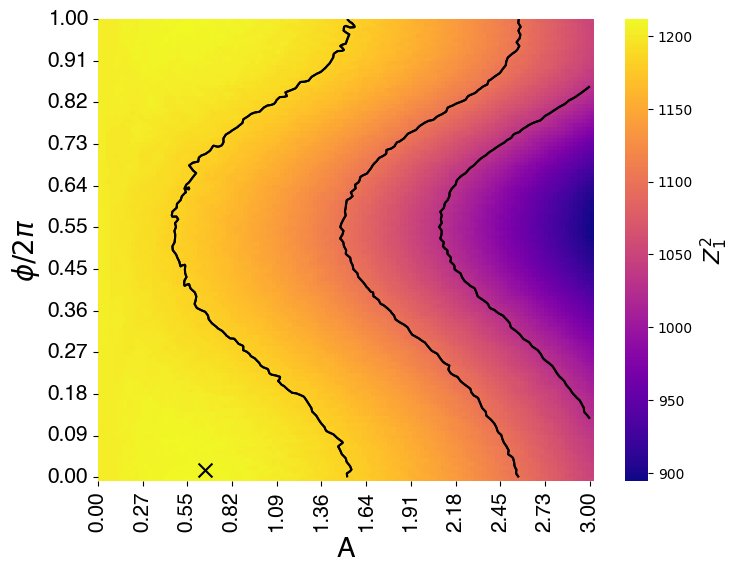}
    \caption{The results of the demodulation analysis on the combined NICER observations. The color-scale representation shows the peak frequency's power change as a function of demodulation amplitude (A) and phase ($\phi$). The highest $Z^2_1$ value, marked by 'x,' indicates the most significant spin period detection, giving the best fit A and $\phi$. The black lines show the 1, 2, and 3-$\sigma$ significance contours.}
    \label{fig:makishimaplot}
\end{figure}

Based on the two \nice\ observations (separated by $\sim$ three months) with 70 ks total exposure in 2021, we attempted to constrain the spin period evolution of \src. The \nice\ data provides the most accurate timing data given its superb resolution ($0.3 < \mu$s). Following \citet{Makishima2021}, we employed a demodulation method to consider the orbital motion. For all photon event arrival times with respect to the reference time 01/01/1998 00:00:00 UTC, we fit $\Delta t = A \sin \left(\frac{2\pi t}{P_{\rm orb}} + \phi_0\right)$ where $P_{\rm orb} = 1.76$ hr (orbital period), $A$ is an amplitude and $\phi_0$ is a phase shift. $A$ and $\phi_0$ are the two unknown parameters we varied to calculate the confidence level contours shown in Figure \ref{fig:makishimaplot}. The demodulation parameters should be constant across all observations. Using both \nice\ observation data, we constrained $A < 1.5$ [s] and found that $\phi$ is unconstrained. Our results suggest that the effects of orbital modulation may be intrinsically small (e.g., possibly due to a face-on view of the orbital plane) or we need more high-resolution X-ray timing data over a longer baseline. Either way, the orbital modulation does not significantly contribute to determining the spin period's uncertainty in the current X-ray data. By adopting $A < 1.5$ [s] and using all \nice\ observation data (with 73 ks total exposure over 104 days), we determined the spin period most accurately ($P = 29.6089\pm0.0002$ s) among the previous timing studies presented in \cite{Oliveira}. We applied the best orbital parameters ($A < 1.5$ [s] for a range of $\phi = 0\rm{-}1$) to \XMM\ pn data in 2019 and 2022. We performed joint $Z^2_2$ tests with the \xmm\ pn and \nice\ timing data spanning over three years, assuming a non-zero spin derivative. We found no significant period difference and derived an upper limit of the spin period derivative of $|\dot{P}| < 1.8\times10^{-12}$ s/s with 90 \% confidence. 

\begin{deluxetable*}{lcccccccc}[ht]
\tablecaption{Summary of Timing Analysis}
\tablecolumns{6}
\tablehead{
\colhead{Instrument}
&
\colhead{{Peak frequency [mHz]}}  
&
\colhead{Peak period [sec]}
&
\colhead{Counts}
&
\colhead{ $Z^2_1$ }
& 
\colhead{False Alarm Probability$^b$} 
&
\colhead{$P $ [\%]$^a$}  
}
\startdata   
\xmm\ 0.3--2 keV & & &  & &  & \\
 \hline
\textbf{0842570101 } \\
MOS & 33.78(3) & $29.61(1)$ & 3890 & 118.82 & $2.5\times 10^{-23}$ & 25(1) \\
pn &33.78(3) & $29.61(2)$ & 6350 & 200.8 & $3.5\times 10^{-41}$ & 27(1)\\
\textbf{0902500101}\\
MOS & 33.77(1) & 29.609(5) & 9142 & 263.37 & $3.1\times 10^{-54}$& 24.3 (9) \\
pn  & 33.77(1) & 29.608(6) & 22,853 & 814.64 & $5.8\times 10^{-174}$ & 27.9 (5) \\
\hline
\hline
\nice\ 0.3--2 keV &&&&& \\
\textbf{459201020*} & 33.773(1) & 29.6095(8) & 87,105 & 989.29 & $1.1\times 10^{-212}$ & 15.5(3) \\
\textbf{459201030*} & 33.772(5) & 29.610(4) & 35,824 & 484.98 & $9.2\times 10^{-104}$ & 16.8(4)\\
\textbf{Joint} & 33.7735(3) & 29.6089(2) & 122,929 & 974.35 & $1.9\times 10^{-209}$ & 14.9(2)\\
\hline
\hline
\nustar\ 3--10 keV &&&&& \\
\hline
\textbf{30801006002}\\
FPMA + FPMB & 33.767(7) & 29.614(6) & 2966 &  37.11 &  $4.7\times 10^{-5}$ & 16.2(1.8) \\
\hline
\enddata
\label{tab:timing}
 All errors shown are at 90\% confidence intervals. The uncertainty on the last digit of peak frequency, peak period, and pulse fraction is shown in parentheses next to the observed value. The results are from the $Z^2_1$ test in the narrow frequency band. The total counts for each observation, instrument, and energy band are also listed.  

$^a$Pulsed fractions [\%] and upper limits were calculated by {\tt HENzsearch}  \citep{matteo_bachetti_2022_6394742}. 

$^b$ The false alarm probability, the probability that the peak signal was generated by noise, was calculated with {\tt Stingray} function \textbf{z2\_n\_probability}.
\end{deluxetable*}

\section{Spectral Analysis} \label{sec:spec}

We jointly fit the \XMM\ and \nice\ spectra with phenomenological models. The \nustar\ observation saw a substantial flux enhancement that coincided with an increase in optical flux. However, it is uncertain if this event represents a dwarf nova outburst (van Dyk et al., in preparation). The unabsorbed X-ray flux was measured to be $F_X = 7.6 \times 10^{-12} \;$ \fluxcgs\ in the 0.3 - 12 keV band, a factor of four greater than all other observations with $F_X \sim 2 \times 10^{-12}\;$ \fluxcgs\ . Thus, we fit its data separately. Various combinations of X-ray absorption and thermal plasma spectral components were applied. Due to the improved photon statistics over the previous X-ray study \citep{Oliveira}, we could perform a phase-resolved X-ray spectral analysis and search for \sout{the} spin variation.

\subsection{Phase-Averaged Spectral Analysis}
    
For the phase-averaged spectral analysis, phenomenological models were fit to the joint \XMM\ and \nice X-ray spectra and \nustar\ X-ray spectra to characterize the X-ray spectral properties further and expand upon the analysis previously presented by \cite{Oliveira}. All spectral fittings were performed using {\tt XSPEC} version 12.13.1  \citep{Arnaud1996}. Each spectral model included {\tt tbabs} to account for ISM absorption using the Wilms abundance data \citep{Wilms2000}. Furthermore, {\tt constant} is included as a cross-normalization factor between different observations in the joint fit. We set the constant factor to 1 for the earliest \xmm\ observation in 2019. 
\begin{deluxetable*}{lccccccc}[ht]
\tablecaption{Phenomenological model fits to X-ray spectra}
\tablecolumns{8}
\tablehead{
\colhead{Parameter}   
&
\colhead{{\tt pow}$^e$}  
& 
\colhead{{\tt APEC}$^e$}
& 
\colhead{$2 \cdot${\tt APEC}}
&
\colhead{{\tt pow} + $2 \cdot${\tt APEC}}
&
\colhead{ {\tt bbody} + $2 \cdot${\tt APEC}}
&
\colhead{$3 \cdot${\tt APEC}}
&
\colhead{$3 \cdot${\tt APEC} + {\tt gauss}}
}
\startdata  
$C_{XMM}^a$  & $0.84 \pm 0.02 $ & 0.83 & $0.83 \pm 0.02$ & 0.83 $\pm 0.02$ & 0.83 $\pm 0.02$ & $0.83 \pm 0.02$ & $0.83 \pm 0.02$\\
$C_{2021}^a$ & $1.09 \pm 0.02$ &  1.10 & $1.06 \pm 0.02$ & $1.02 \pm 0.02$ & $1.03 \pm 0.02$ & $1.03 \pm 0.02$ & $1.03 \pm 0.02$\\
$C_{2022}^a$ & $1.23 \pm 0.03$ &  1.25 & $1.19 \pm 0.03$ & $1.15\pm 0.03$  & $1.16 \pm 0.03$ & $1.16 \pm 0.03$ & $1.16 \pm 0.03$\\
$N^{(i)}_H (10^{20} \rm{cm}^{-2})^b$ & $15.5 \pm 0.5$ & 4.99 & $1.4 \pm 0.3$ & 9.2 $\pm 0.9$ & 6.2 $\pm 1.0$ & $4.2_{-0.5}^{+0.6}$ & $4.2 \pm 0.6$\\
$\Gamma$ & $2.44 \pm 0.04$ & ... & ... & 1.6 $\pm 0.1$ & ... &  ... & ...\\
$kT_1$ (keV) & ... & 2.13 & $0.85\pm 0.01$ & $0.30_{-0.01}^{+0.02}$  & 0.18$\pm 0.01$ & $0.30\pm0.01$ &  $0.33\pm0.01$\\ 
$kT_2$ (keV) & ... & ... & $4.39_{-0.18}^{+0.16}$  &  $1.12\pm0.02$ & $0.94\pm0.02$ & $1.01_{-0.02}^{+0.03}$ & $1.01_{-0.02}^{+0.03}$\\
$kT_3$ (keV) & ... & ... & ... & ... & $6.35_{-0.50}^{+0.73}$ & $4.90_{-0.23}^{+0.22}$ & $4.86\pm0.23$ \\
$Z^c (Z_\odot)$ & ... & 0.07 & $0.51\pm 0.06$ & $0.13_{-0.03}^{+0.08}$ & $0.57_{-0.09}^{+0.10}$ & $0.53_{-0.06}^{+0.07}$ & $0.52_{-0.06}^{+0.07}$ \\
$E_{\rm line}$ (keV) & ... & ...  & ... & ... & ... & ... & 6.4 \\
$\sigma_{\rm line}$ (keV) & ... & ... & ... & ... & ... & ... & 0.01 \\
$EW_{\rm line}$ (eV) & ... & ... & ... & ... & ... & ... & $<73$\\
$F_X (10^{-12} \frac{\mathrm{erg}}{\mathrm{cm}^2 \mathrm{s}})^d$ & 2.51& 3.36 & 6.72 & 8.19 & 7.72& 7.04 & 7.09 \\ 
$\chi^2_{\nu}$ (dof) & 1.93 (2330) & 2.18 (2329) & 1.16 (2327) & 1.06 (2325) & 1.06 (2325) & 1.01 (2325) & 1.01 (2324) \\
\enddata
\label{tab:wonu_prelimfits}
All errors shown are $90 \%$ confidence intervals. \\
$^a$ Cross-normalization factors of the \xmm\ data in 2022 ($C_{XMM}$), and the \nice\ data in 2021 ($C_{2021}$) and 2022 ($C_{2022}$) with respect to the \xmm\ observation in 2019. \\
$^b$ The ISM hydrogen column density associated with {\tt tbabs} which is multiplied to all the models. \\
$^c$ Abundance relative to solar. \\
$^d$ 3 -- 10 keV flux of the \XMM\ pn (2019) data. \\
$^e$ Reduced $\chi^2$ is too great to calculate error. \\
$^*$ The parameter is frozen. \\
\end{deluxetable*}

\begin{deluxetable*}{lccc}[ht]
\tablecaption{Phenomenological Model Fits to the \nustar\ spectra}
\tablecolumns{4}
\tablehead{
\colhead{Parameter}   
&
\colhead{{\tt pow}$^e$}  
& 
\colhead{{\tt APEC}$^e$}
& 
\colhead{{\tt APEC} + {\tt gauss}}
}
\startdata  
    $N^{(i)}_H (10^{20} \rm{cm}^{-2})^a\;*$ & $4.2$ & 4.2 & 4.2\\
$\Gamma$ & $2.27 \pm 0.06$ & ... & ... \\
$kT_1$ (keV) & ... & $8.2_{-0.5}^{+0.6}$ &  $8.4_{-0.6}^{+0.7}$ \\
$Z^b (Z_\odot)$ & ... & $0.40_{-0.11}^{+0.13}$ & $0.32_{-0.12}^{+0.13}$ \\
$E_{line}$ (keV)* & ... & ... & 6.4 \\
$\sigma_{line}$ (keV)* & ... & ... &  0.01 \\
$EW_{line}$ (eV) & ... & ... & $174_{-76}^{+75}$ \\
$F_X (10^{-12} \frac{\mathrm{erg}}{\mathrm{cm}^2 \mathrm{s}})^c$  & 4.21 & 4.39 & 4.39\\
$\chi^2_{\nu}$ (dof) & 1.36 (256) & 1.05 (255) & 0.99 (254)\\
\enddata
\label{tab:nustar_fits}
All errors shown are $90 \%$ confidence intervals. \\
$^a$ The ISM hydrogen column density associated with {\tt tbabs}, which is multiplied to all the models. \\
$^b$ Abundance relative to solar. \\
$^c$ 3 -- 10 keV  flux of the \nustar\ data. \\
$^*$ The parameter is frozen. \\
\end{deluxetable*}

Fitting an absorbed power-law model to the joint spectrum resulted in a poor fit to the data ($\chi^2_\nu = 1.93$), largely due to atomic lines at 6.7--7 keV, indicating that the X-ray emission is thermal. We applied an emission spectrum model ({\tt APEC}) that accounts for the emissions from collisionally-ionized diffuse gas due to accretion. A single temperature {\tt APEC} model did not fit the data well ($\chi^2_\nu = 2.18$) with the best-fit temperature of $kT = 3.37$ keV. We then fit an absorbed 2-temperature model ({\tt APEC}+{\tt APEC}). The fit was significantly improved ($\chi^2_\nu = 1.16$) with the best-fit temperatures $kT = 0.85$ and 4.39 keV. Despite the improvement, the model failed to fit the X-ray spectra above 6--7 keV accurately. We added a third {\tt APEC} model with a higher plasma temperature. The 3-temperature model ({\tt APEC}+{\tt APEC}+{\tt APEC}) yielded a better fit with $\chi^2_\nu = 1.01$ with the best-fit temperatures $kT = 0.30, 1.01$ and 4.90 keV. Some residuals are present below 2 keV, possibly due to complex X-ray absorption often seen in IPs \citep{Mukai2021}. The abundance ($Z$) was linked between the different {\tt APEC} components and we found $Z = 0.54_{-0.06}^{+0.07} \;Z_{\odot}$ for the 3-T model. 

We also considered a combination of thermal and power-law components (e.g., {\tt APEC+APEC+PL}). However, the fit was not improved ($\chi^2_\nu = 1.06$) compared to the thermal-only models. Thus, we conclude that there is no evidence of non-thermal X-ray emission based on the broadband X-ray data. Although the lower-energy residuals improved, the fit was again poor in the region of the hard X-ray photons. We also replaced the lowest temperature {\tt APEC} model with a blackbody model by fitting an absorbed {\tt BBODY+APEC+APEC} model. The fit quality was again similar to the thermal and power-law model ($\chi^2_\nu = 1.06$) and the best-fit blackbody temperature was 0.18 keV. The soft blackbody component could originate from the base of the accretion column as observed from other mCVs \citep{Ramsay_2002}.  
    
In IPs, some X-rays emitted from the accretion column can be reflected by the WD surface or absorbed by the accretion curtain. These effects cause a neutral Fe K-$\alpha$ fluorescence line at 6.4 keV and complex X-ray absorption at lower energies \citep{Mukai2021}. Following previous X-ray studies (e.g., \citet{Hailey2016}), we added a Gaussian line component at $E=6.4$ keV to the absorbed 3-T {\tt APEC} model to account for the neutral Fe K-$\alpha$ fluorescence line. We fixed the line energy and set the line width to $\sigma = 0.01$ keV. The fit improved by only $\Delta\chi^2 = 1.3$ over 2325, indicating that adding the Gaussian component is not statistically significant.  We found a 3--$\sigma$ upper limit of the Gaussian component to EW $< 79$ eV. The other spectral parameters remained nearly unchanged from the fit without the Gaussian line component. From these fits, we constrained $n_H$ to $4.2 \pm 0.6 \times 10^{20}\; \rm{cm}^{-2}$. The best-fit absorbed flux was $7 \times 10^{-13}\;$\fluxcgs (Table \ref{tab:wonu_prelimfits}). 

As the X-ray spectrum of \src\ extends down to 0.3 keV, X-ray absorption seems insignificant. We added a partial covering absorption component ({\tt pcfabs}), which accounts for X-ray absorption in the accretion curtain for IPs \citep{Hailey2016}. However, it did not improve the spectral fits in any of the aforementioned cases. 

For the \nustar\ spectrum, we froze $n_H = 4.2 \times 10^{20}\; \rm{cm}^{-2}$ as the lack of counts left the value otherwise unconstrained. Here, the absorbed power-law model also fit poorly due to the atomic lines in the 6.7 - 7 keV region  ($\chi^2_\nu = 1.36$). A single temperature {\tt APEC} model improved the fit to $\chi^2_\nu = 1.05$. The highest temperature component of the \xmm\ and \nice\ fits (formerly 4.9 keV) is better characterized by a temperature of 8.4 keV in the broader 3 - 30 keV energy band. Adding a Gaussian component to account for the neutral Fe line further improved the fit to $\chi^2_\nu = 0.99$, increasing temperature to 8.4 keV and decreasing the abundance to $Z = 0.32_{-0.12}^{-0.13} M_{\odot}$. The EW of the Gaussian component ($174_{-76}^{+85}$ eV) was substantially stronger here than in the joint fitting, which weakly indicated the presence of an Fe line. A partial covering absorption component, however, similarly did not improve the fit to the $\nustar$ observation. Finally, we fit best-fit $3 \cdot${\tt APEC} + {\tt gauss} model from the \xmm\ and \nice\ observations, freezing all components of the model except for normalization factors. The fit was poor with $\chi^2_\nu = 1.66$ for 254 d.o.f. with prominent residuals in the above 7 keV, confirming that the X-ray spectrum has hardened during the outburst that \nustar\ observed.

\begin{figure}[ht]
\begin{center} 
\includegraphics[width=8.8cm]{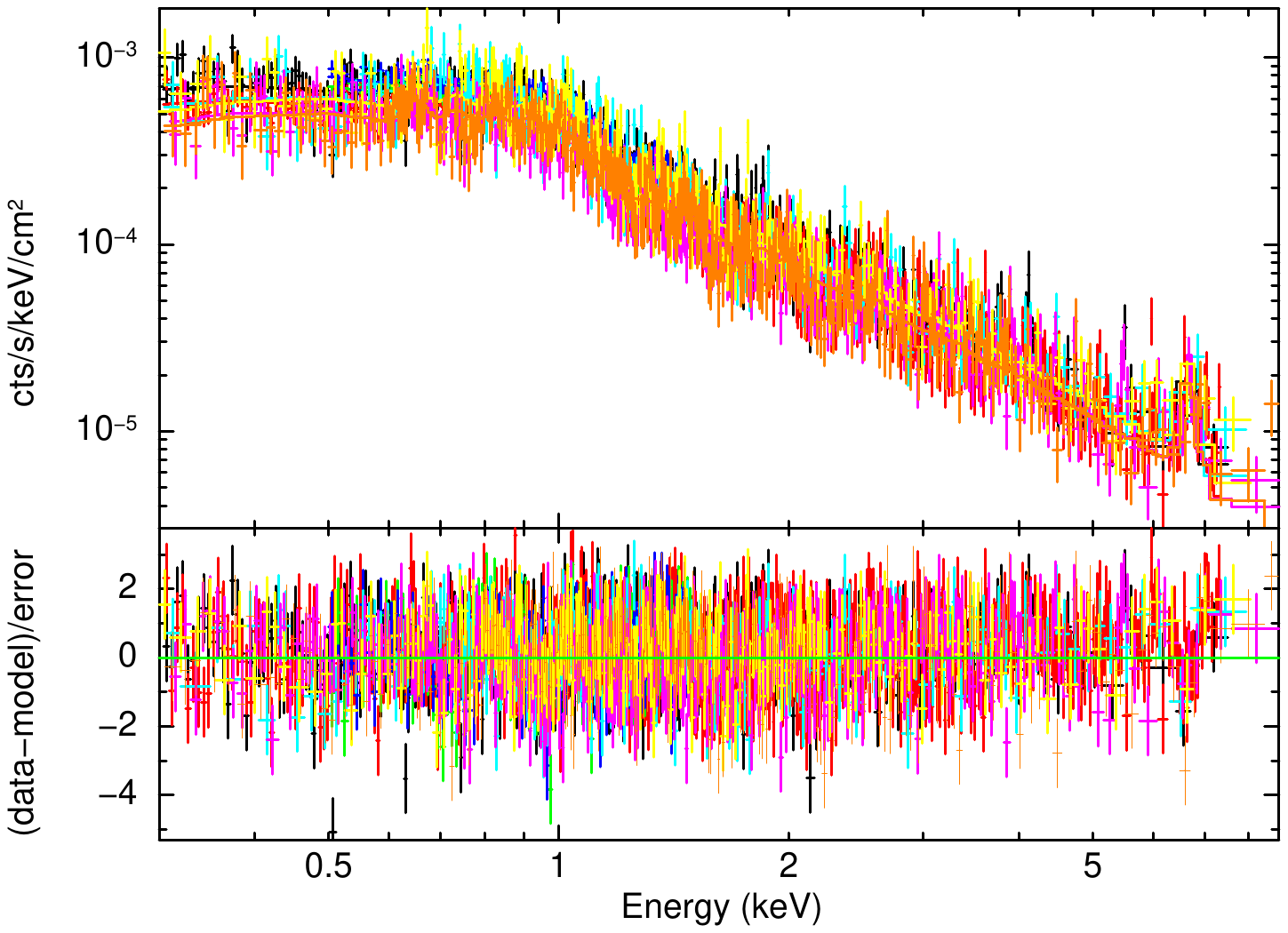}
\includegraphics[width=8.8cm]{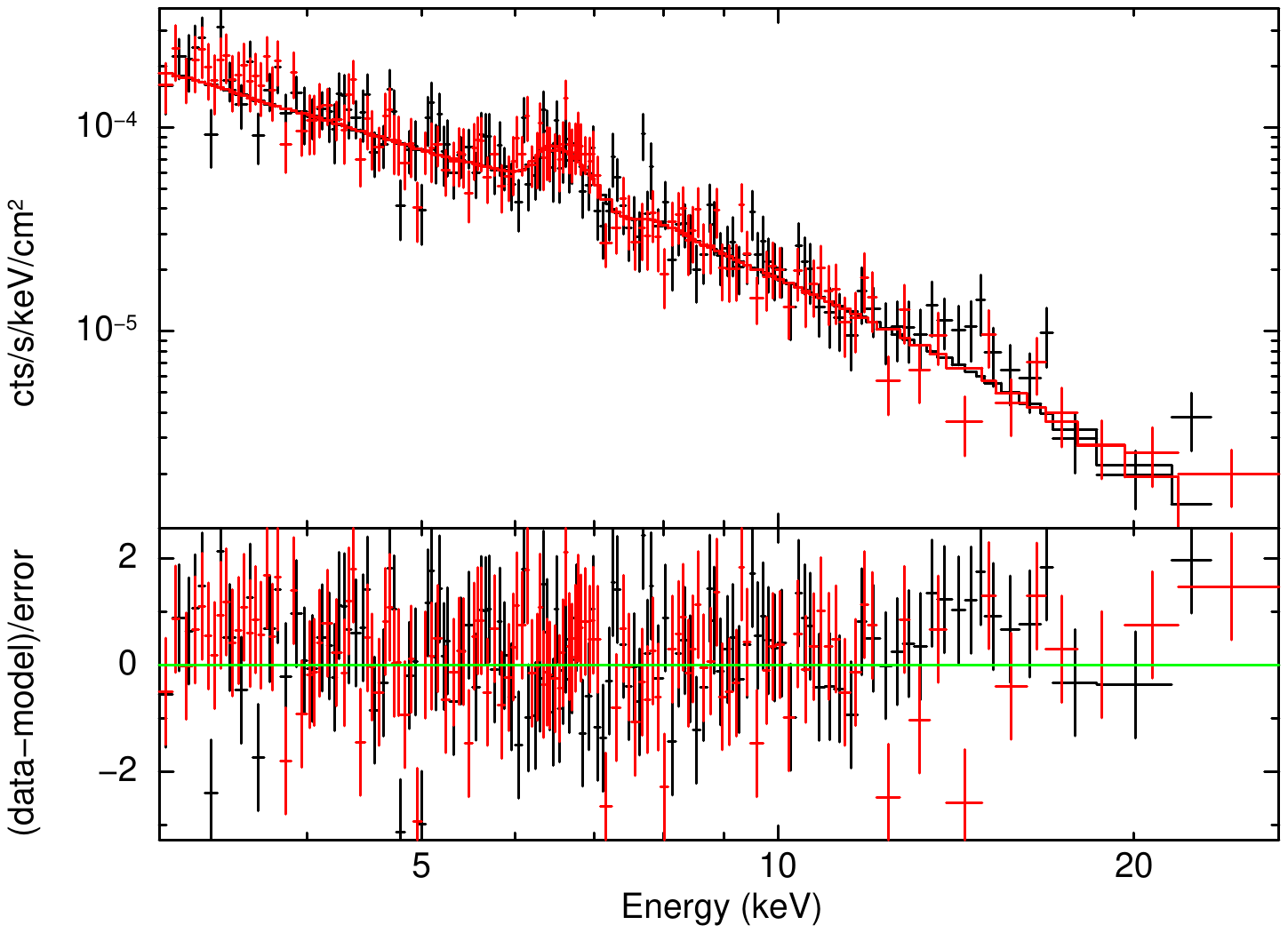}

\end{center} 
\caption{ Left: \xmm\, \nice\ spectra and residuals of \src\ fit with an  {\tt tbabs}*({\tt APEC}+{\tt APEC}+{\tt APEC} + {\tt gauss}) model, which yielded a $\chi^2_{\nu} = 1.01$ for 2324 degrees of freedom. The 2019 \xmm\ observation (obsID = 0842570101) is presented in the 0.3 - 10 keV band with MOS1 in cyan, MOS2 in yellow, and pn in black. Similarly, 2022 \xmm\ observation (obsID = 0902500101) is presented in the 0.3 - 10 keV band with MOS1 in magenta, MOS2 in orange, and pn in red. The \nice\ observations are presented in the 0.5 - 1.5 keV band, with obsID = 459201020* in green and obsID = 459201030* in blue. Right: \nustar\ spectrum and residuals fit with a {\tt tbabs}*{\tt APEC}, which yielded a $\chi^2_{\nu} = 0.99$ for 254 degrees of freedom. The \nustar\ observation is presented in the 3-30 keV band, with FPMA in black and FPMB in red.}
\end{figure}

\subsection{Phase-Resolved Spectral Analysis}

\begin{figure}[ht]
    \centering
    \begin{minipage}{0.48\linewidth}
        \includegraphics[width=\linewidth]{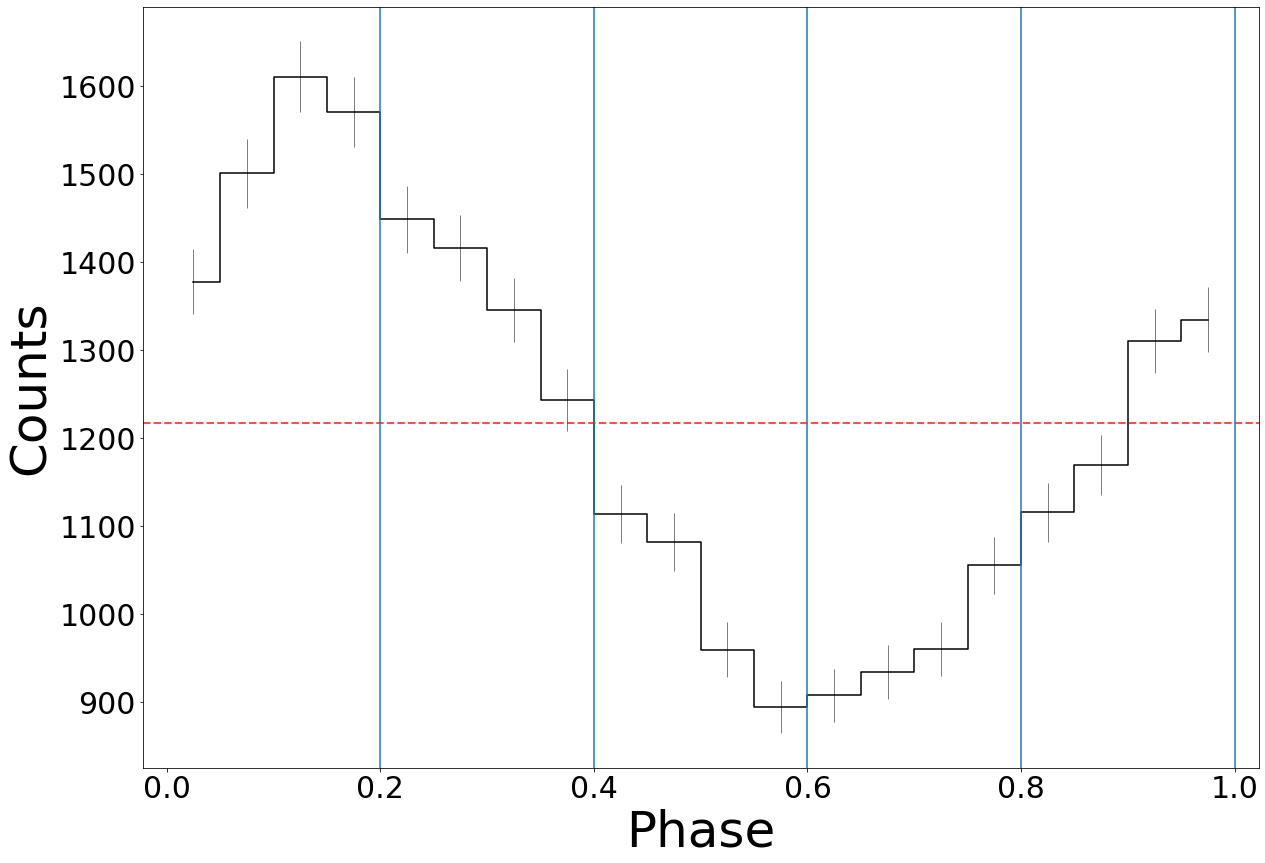}
    \end{minipage}
    \hfill
    \begin{minipage}{0.48\linewidth}
        \includegraphics[width=\linewidth]{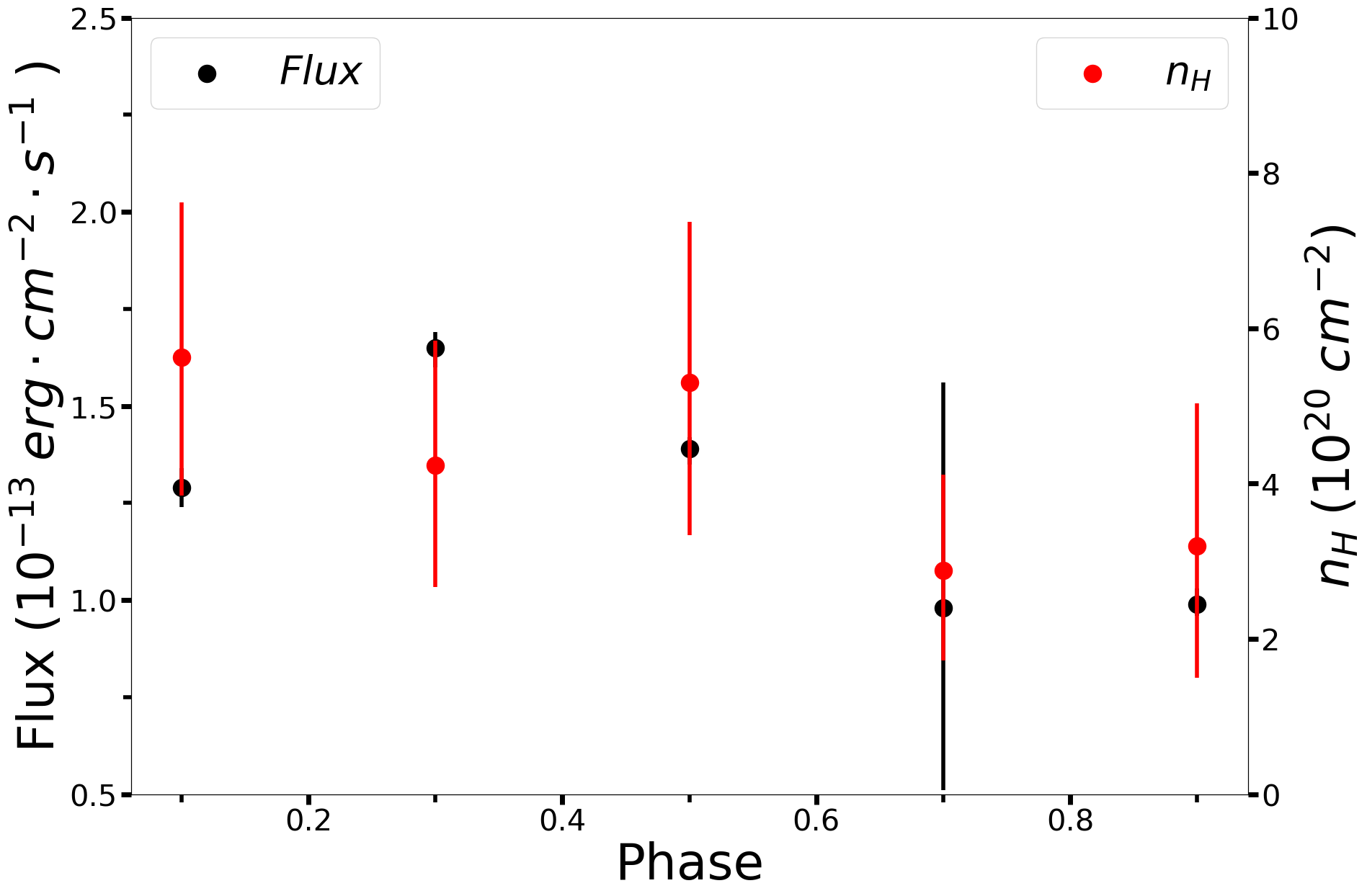}
    \end{minipage}

    \begin{minipage}{0.48\linewidth}
        \includegraphics[width=\linewidth]{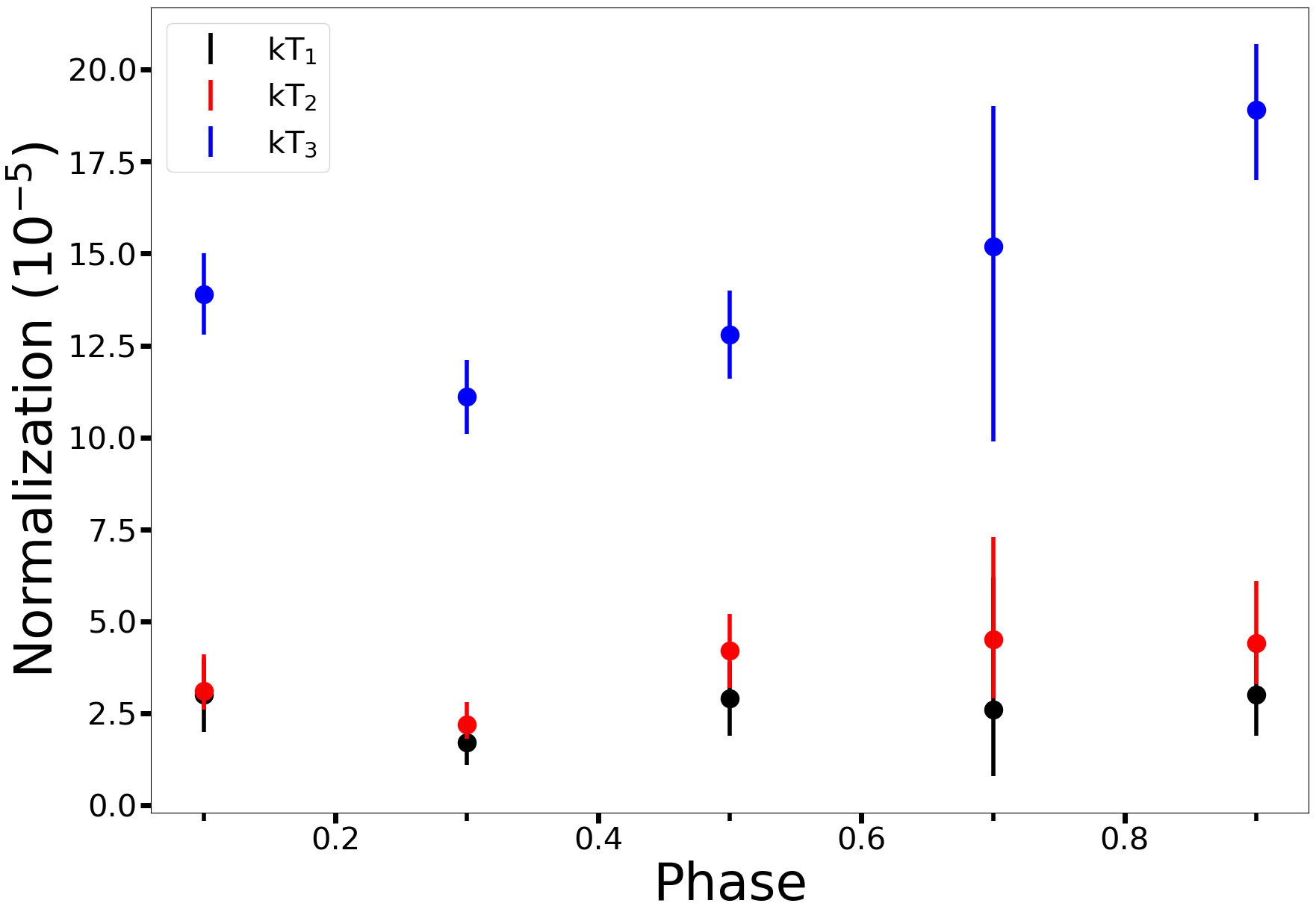}
    \end{minipage}
    \hfill
    \begin{minipage}{0.48\linewidth}
        \includegraphics[width=\linewidth]{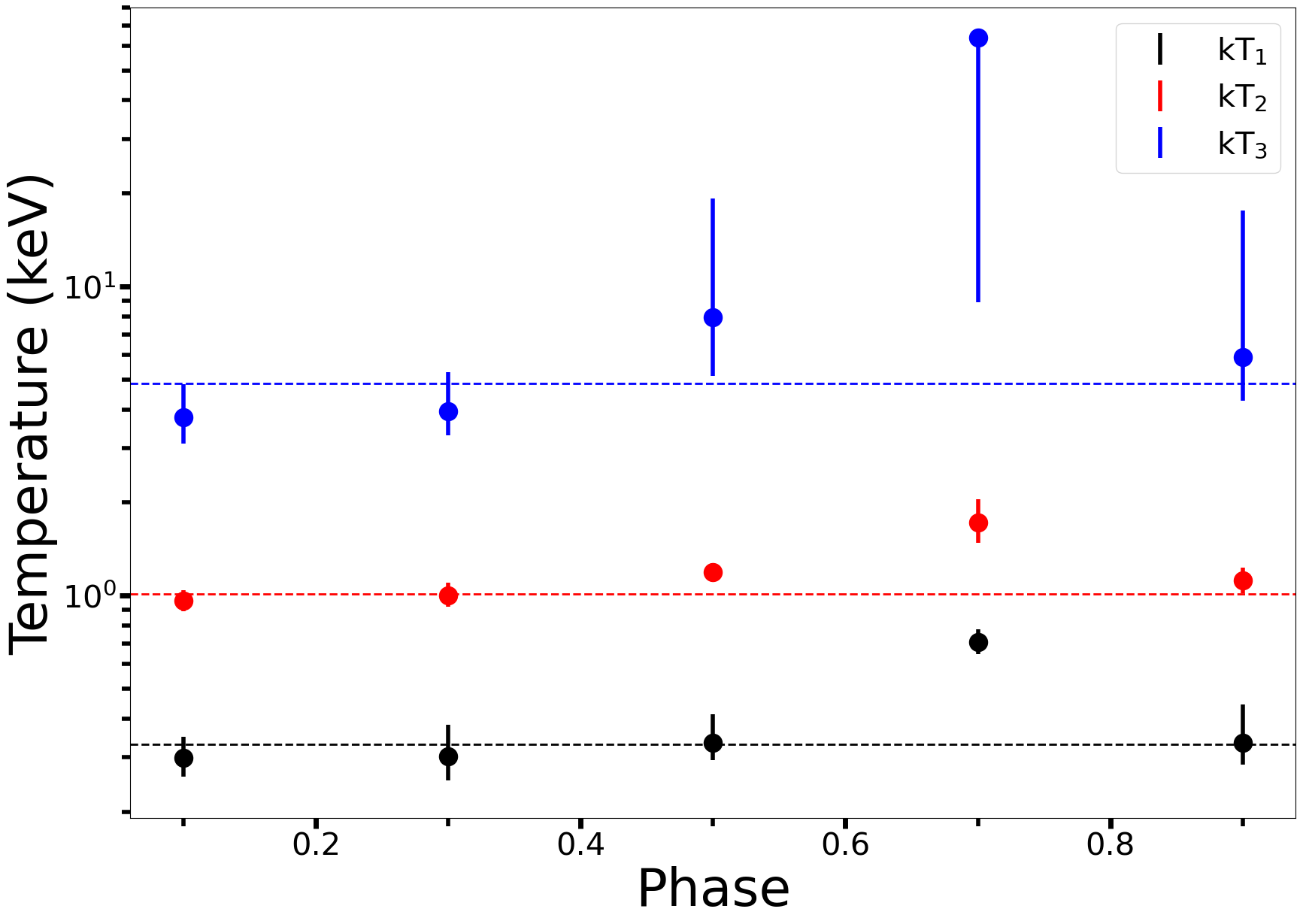}
    \end{minipage}

    \caption{{\it Top Left:} The 2022 EPIC-pn light curve folded at the peak spin period, demonstrating the phase bins the spectra were cut into. The dotted red line denotes the mean counts, while the blue lines mark the 5 phase intervals ($\phi = 0.0\rm{-}0.2, 0.2\rm{-}0.4, 0.4\rm{-}0.6, 0.6\rm{-}0.8$ and 0.8--1.0), where \xmm\ spectra were separately fitted. {\it Top Right:} The soft X-ray (0.3 - 2 keV) flux plotted against the fitted $n_{H}$ across the phase bins with all temperatures fit freely. It can be observed that fitted absorption decreases in the brightest bins. {\it Bottom:} The flux normalization variation and plasma temperature variation for each {\tt APEC} component across five phase intervals. For comparison, the phase-averaged values are shown as dotted lines. The phase-resolved spectroscopy demonstrated a change in normalization that similarly affects all three single temperature components of the model. The fourth phase bin at $\phi = 0.6\rm{-}0.8$ shows a substantial, but poorly constrained, spectral hardening that is pronounced across all temperature components.}
    \label{fig:phasecuts}
\end{figure}
    
We performed phase-resolved spectral analysis using the best-fit spin period. As IPs often demonstrate energy-dependent spin modulation, \src\ may demonstrate some phase variation in its X-ray spectra (\cite{1988Rosen}, \cite{Joshi2022}). 
We calculated phase values for all source photon events based on each observation's best-fit spin period ($P= 29.61$ s). We extracted X-ray spectra from five phase intervals with equal length ($\Delta\phi = 0.2$; Figure \ref{fig:phasecuts} left panel). The \nice\ data was excluded from the phase-resolved spectral analysis due to the lack of source counts in the energy band above 1.5 keV. We also exclude the \nustar\ data as it is described best by a different model than the other observations. We refrain from presenting its own phase-resolved spectroscopy due to insufficient counts (a few hundred per phase bin) and the unconstrained $n_H$ in this energy band.

We fit each phase-resolved spectrum with an absorbed 3-temperature model {\tt constant}*{\tt tbabs}*({\tt APEC}+{\tt APEC}+{\tt APEC}), varying all parameters, except for the abundance which was frozen to the phase-averaged best-fit abundance. The fit quality was excellent for the first three bins with $\chi^2_\nu$ between 0.95 and 1.01. However, phase bins 4 and 5 contained the least bins and worst fits ($\frac{\chi_\nu^2}{\nu} = \frac{0.91}{401}$ and $\frac{\chi_\nu^2}{\nu} = \frac{0.89}{434}$). This also led to poor constraints of the fit parameters, particularly the highest temperature component, in phase bin 4. The source undergoes spectral hardening at its spin-phase flux minima (in phase bins 3 and 4), particularly in its lower temperature components, as demonstrated by the increase in $kT_1$ and $kT_2$, while $kT_3$ is unconstrained. We also present the flux variability in the 0.3--2 keV range, where the spin-period signal was significantly detected. Again, the soft component of the X-ray spectrum reaches a maximum at the flux maximum. This result confirms the behavior demonstrated by the hardness ratio modulation. Meanwhile, the X-ray absorption, quantified by $n_H$, shows a weak anti-correlation with soft X-ray flux. To avoid degeneracy between flux normalization and the fitted temperature of each component, we compared the relationship between flux and $n_H$ with frozen $kT_1$, $kT_2$, and $kT_3$ but found the same behavior.

\section{White Dwarf Mass Measurement} 
\label{sec:mass}
In general, X-ray emission in CVs is powered by mass accretion, converting the gravitational potential energy of incoming particles to X-rays by radiative cooling via the shock-heated gas flow. 
In non-magnetic CVs (nmCVs), the accretion disk tends to approach the WD surface, where a boundary layer is formed and dissipates most of the gravitational energy into X-ray emission.  
In magnetic CVs (mCVs), accreting material is funneled onto the WD poles along the magnetic field lines. It forms a column of infalling gas, which is first heated by a standoff shock and then cools toward the WD surface by producing thermal X-ray emission \citep{Mukai2017}. In either case, X-ray spectra exhibit multi-temperature components, either due to the boundary layer or accretion flow with varying plasma temperature and density profiles. X-ray emission from nmCVs and mCVs has been well studied, both observationally and theoretically, and can be used to determine WD mass (e.g., \citet{Shaw2020}). While \citet{Oliveira} suggested \src\ is an IP and thus its X-ray emission results from a tall accretion column, the detection of dwarf novae could indicate that the WD may be weakly magnetized like nmCVs. Given that the nature of \src\ is unknown, we considered both the nmCV and mCV cases to determine the WD mass below. In both approaches, we jointly fit all phase-averaged X-ray spectra from \XMM\ and \nustar\ observations. Although \citet{Oliveira} and our results found no significant X-ray absorption at low energies, following \citet{Hayashi2021}, we exclude data below 3 keV to prevent any complex absorption from affecting our following WD mass measurement. \nice\ data is excluded as it is dominated by background contamination above 1.5 keV. We also freeze $N_H = 4.2 \times 10^{20} \;\mathrm{cm}^{-2}$ in {\tt tbabs}, according to our best phenomenological fit, as it does not contribute to the spectral shape above 3 keV. 

\subsection{Non-magnetic CV case} 
\begin{figure} 
    \begin{center}
    \includegraphics[width=0.60\textwidth]{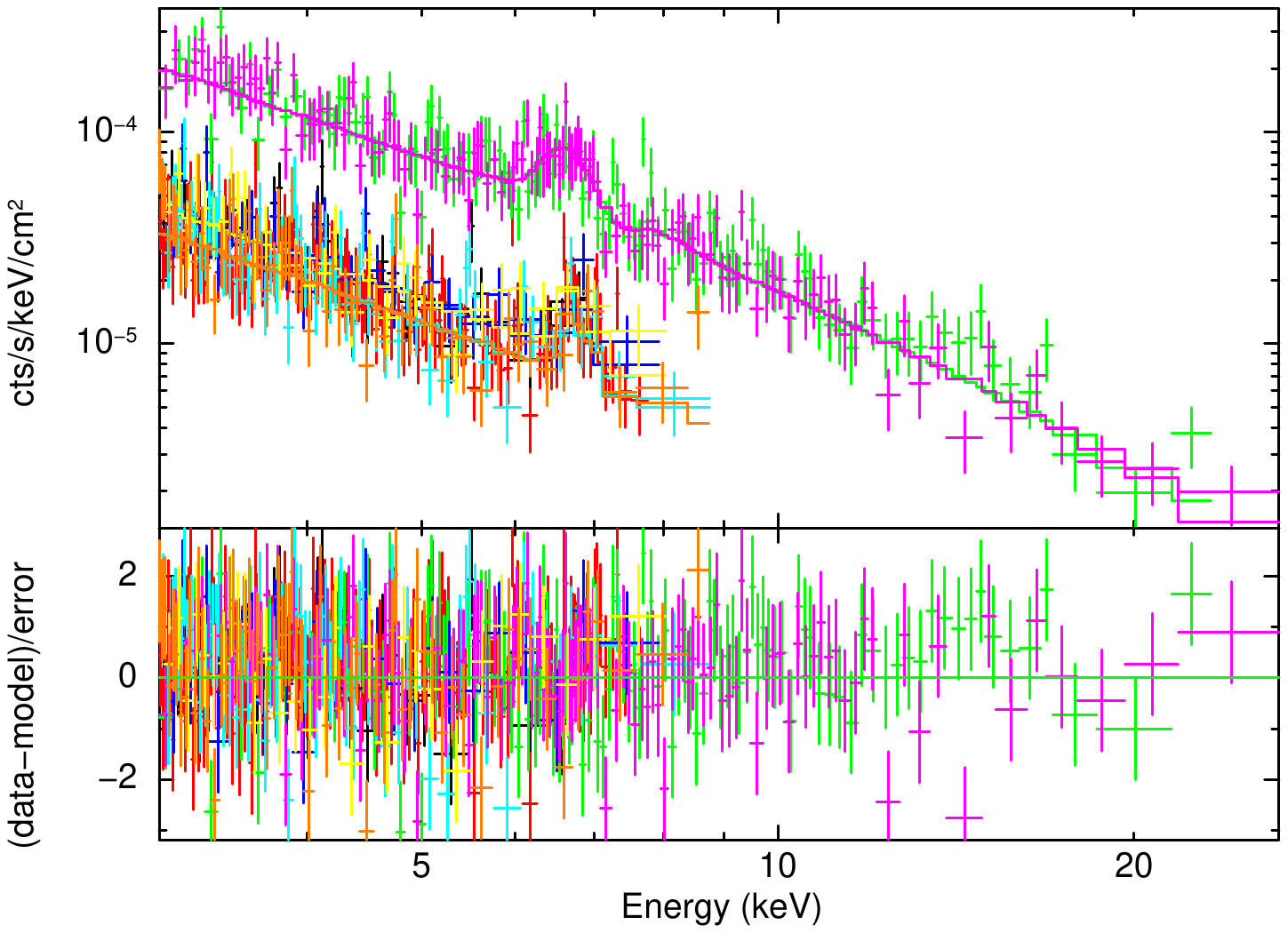} 
    \caption{The \xmm\ and \nustar\ spectra and residuals of \src\ fit with {\tt tbabs*(mkcflow+gauss)} with $\chi_\nu^2 = 0.98$ for 628 degrees of freedom. The 2019 \xmm\ observation (obsID = 0842570101) is presented in the 0.3 - 10 keV band with MOS1 in blue, MOS2 in yellow, and pn in black. Similarly, 2022 \xmm\ observation (obsID = 0902500101) is presented in the 0.3 - 10 keV band with MOS1 in cyan, MOS2 in orange, and pn in red. The \nustar\ observation is presented in the 3-30 keV band, with FPMA in black and FPMB in red.}
    
    \label{fig:nmCVfit} 
    \end{center}
    \end{figure}
    
In nmCVs, X-ray photons are emitted from a boundary layer between the accretion disk and the WD surface. An absorbed plasma cooling model generally describes thermal X-ray emission from the shock-heated boundary layer (e.g., {\tt mkcflow}). The maximum temperature ($kT_{\rm max}$) in the {\tt mkcflow} model can be attributed to the shock temperature, $T_s$, in the boundary layer. Assuming the gravitational potential energy is converted to heat in the strong shock regime at the inner accretion disk spinning at the Keplerian velocity, the maximum temperature of the disk is given by $k T_{\rm max} = \frac{3}{16} \mu m_{\rm H} \frac{GM}{R}$ where $\mu$ is the mean molecular weight, $m_{\rm H}$ is the mass of a hydrogen atom and $R$ is the WD radius \citep{2002apa..book.....F}. More specifically, \citet{Yu2018} derived an empirical relation $k T_{\rm max} = \frac{3}{16} \alpha \mu m_{\rm H} \frac{GM}{R}$, where $\alpha$ was fit to $0.646\pm0.069$, based on the maximum plasma temperature and known WD mass data of 11 nmCVs. 

Following \citet{Yu2018}, we fit an absorbed cooling flow model with a Gaussian line component to account for the neutral Fe emission line at 6.4 keV. Our spectral model, {\tt tbabs*(mkcflow+gauss)}, fits the \xmm\ and \nustar\ data well, yielding $kT_{\rm max} = 16.2_{-1.8}^{+1.9}$ keV, as shown in Table \ref{tab:mkcflow}. We exclude the {\tt APEC} component used in \citet{Oliveira} since its contribution to the 3 - 30 keV spectra is negligible due to the low temperature ($kT=0.79$ keV). To account for the difference in flux between the \nustar\ observation and the other spectra, we fit the normalizations of {\tt mkcflow} and the Gaussian component separately. As the remaining parameters only depend on fundamental properties of the WD (such as mass), which do not change with flux variations due to increased mass flow and accretion, we fit them jointly across all observations. By considering the uncertainties associated with $kT_{\rm max}$ and $\alpha$, we determined the WD mass to be $M = (0.84\rm{-}1.03) \;M_\odot$. The normalization of the {\tt mkcflow} yields the total mass accretion rates of $\dot{M} = (3.4\pm0.3)\times10^{14},\; (2.6\pm0.3)\times10^{14}\; \rm{and}\; 17.6_{-2.1}^{+2.5} \times10^{14}$ g\,s$^{-1}$ for the 2019 \xmm, 2022 \xmm\ and 2022 \nustar\ observations, respectively.  

\begin{deluxetable*}{lc}[!h]
\tablecaption{The nmCV model fit to the X-ray spectra}
\label{tab:mkcflow}
\tablecolumns{2}
\tablehead{
\colhead{Parameter}   
&
\colhead{{\tt MKCFLOW}}  
}
\startdata  
$N^{(i)}_H (10^{20} \rm{cm}^{-2})^a\;*$ &4.2 \\
$Z^b (Z_\odot)$ & $0.50_{-0.10}^{+0.12}$ \\
$kT_{\rm max}$ (keV) & $16.3_{-1.9}^{+1.8}$ \\
$EW^c_{\rm line}$ (eV) & $12_{-12}^{+37}$ \\
$EW^d_{\rm line}$ (eV) & $138_{-65}^{+70}$ \\
$F_X (10^{-12}$ \fluxcgs)$^e$  & 6.61\\
$F_X (10^{-13}$ \fluxcgs)$^f$  & 35.1\\
$\chi^2_{\nu}$ (dof) & 0.98 (628) \\
\enddata
\label{tab:nustar_mass_fits}
All errors shown are $90 \%$ confidence intervals. \\
$^a$ The ISM hydrogen column density is associated with {\tt tbabs}, which all the models are multiplied by. \\
$^b$ Abundance relative to solar. \\
$^c$ The equivalent width of the Gaussian component with $E =  6.4$ keV and $\sigma = 0.01$ keV for the both \xmm\ datasets.\\
$^d$ The equivalent width of the Gaussian component with $E =  6.4$ keV and $\sigma = 0.01$ keV for the \nustar\ data. \\
$^e$ 3 -- 10 keV flux of the \xmm\ 2019 data. \\
$^f$ 3 -- 10 keV flux of the \nustar\ 2021 data. \\
$^*$ The parameter is frozen. \\

\end{deluxetable*}

\subsection{Magnetic CV case} 
\label{sec:mcv}

Following \citet{Vermette2023}, we employed the {\tt MCVSPEC} model by considering the magnetically confined accretion flow, its plasma temperature and density profiles, and varying X-ray emissivity along the accretion column. The input parameters for {\tt MCVSPEC} are $M$, $\dot{m}$ (specific accretion rate [g\,cm$^{-2}$\,s$^{-1}$]), $R_{m}/R$ where $R_{m}$ is the magnetospheric radius, $Z$ (abundance relative to the solar value) and flux normalization. Our model assumes that the free-falling gas gains kinetic energy from the magnetospheric radius to the shock height ($h_s$). Thus, at the shock height, the free-falling velocity is given by $v_{ff} = \sqrt{2GM (\frac{1}{R+h_s} - \frac{1}{R_m})}$. In IPs, the accreting gas is heated by the strong shock and continues falling toward the WD surface while cooling radiatively by emitting thermal X-rays. It is essential to include the effects of a finite magnetospheric radius and shock height into the model to correctly determine the shock temperature and the WD mass as related to each other through $kT_{\rm s} = \frac{3}{8} \mu m_{\rm H} v^2_{ff}$ \citep{Hayashi2014, Suleimanov2016,  Hailey2016, Shaw2020}. Some X-rays may be reflected off the WD surface or pre-shock region in the accretion curtain. These reprocessed X-ray photons will appear as a fluorescent neutral Fe line at 6.4 keV and a Compton scattering hump above $\sim$ 10 keV. Below, we considered all these effects in our model and fit the broadband X-ray spectra obtained by \xmm\ and \nustar\ . 

We estimate the magnetospheric radius using the X-ray timing results. Since we did not detect a spectral break associated with $R_m$ in the PDS (as suggested by \citet{Suleimanov2016}), we assumed the WD was in spin equilibrium and its spin period was equal to the co-rotation period. To ensure the mass fittings were consistent with spin equilibrium, we linked the $R_m/R$ parameter to mass in {\tt XSPEC} so that its value would adapt to the mass fitting according to  $R_m = \left(\frac{G M P^2}{4\pi^2}\right)^{1/3}$, where $P$ is the spin period \citep{Suleimanov2016}.

\begin{figure} 
    \begin{center}
    \includegraphics[width=0.45\textwidth]{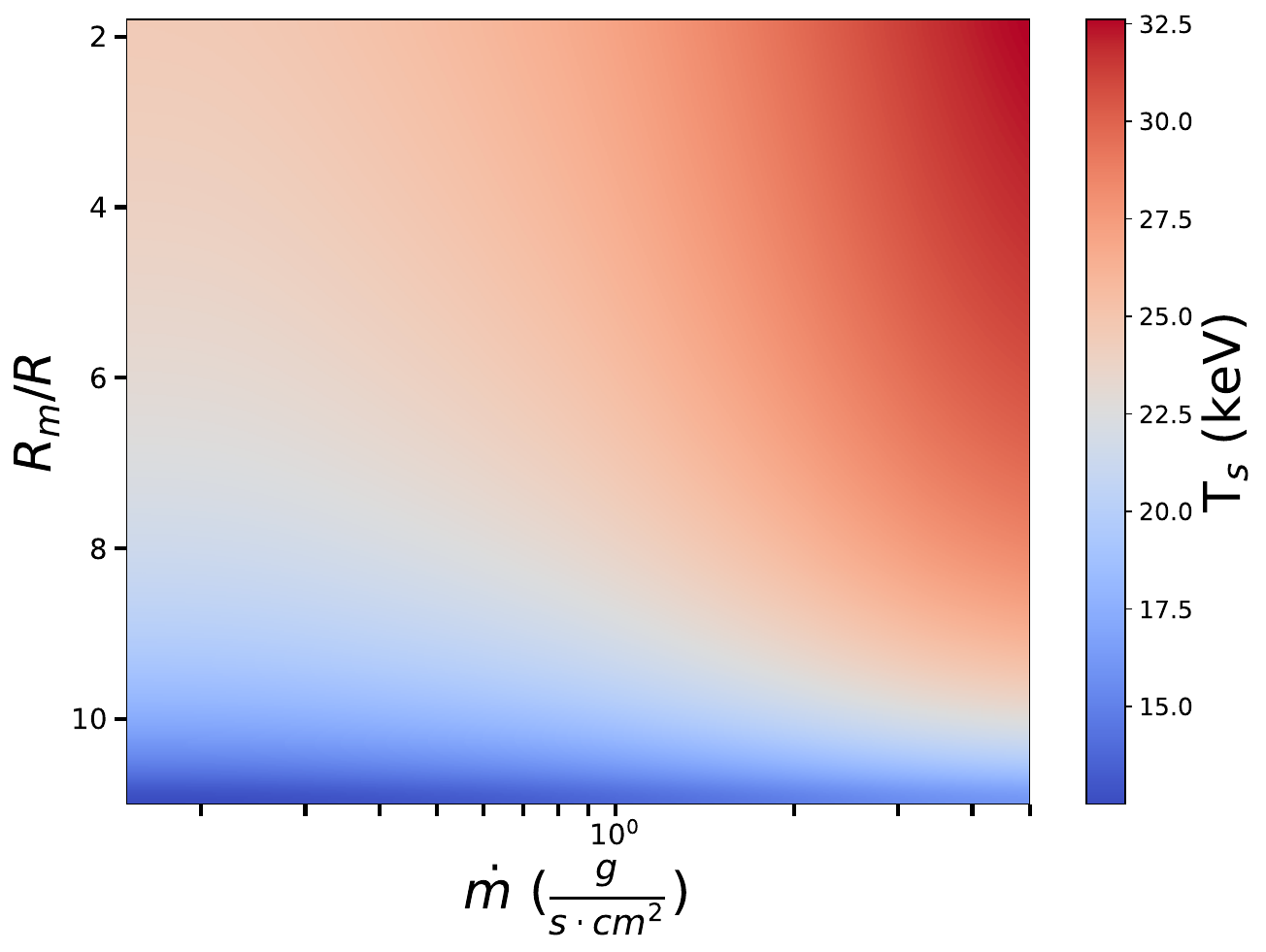} 
    \caption{The variation of shock temperature for an IP with $M = 0.8 M_{\odot}$ as a function of $R_m/R$ and $\dot{m}$. For low $\dot{m}$ IPs (bottom left), which may be representative of the LLIPs and \src, $k T_s$ is substantially lower than in high $\dot{m}$ systems (top right). 
    }
    \label{fig:RmR} 
    \end{center}
\end{figure}

Our WD mass measurement of \src\ is primarily limited by the systematic errors due to the unknown $\dot{m}$. In agreement with previous studies, the unabsorbed X-ray flux, which is used to calculate X-ray luminosity, remains $F_X = 2.1 \times 10^{-12}$ \fluxcgs\ for the 0.3 - 12 keV band in all observations except for the \nustar\ observation where $F_x = 7.6 \times 10^{-12}$ \fluxcgs\ (an increase of $\sim$ 4). By considering the range of possible masses and ratios of magnetosphere radius to WD radius as well as a source distance of 261 pc, we find the mass accretion rate to be $\dot{M} = (0.23\rm{-}9.7)\times 10^{14}  $ [g\,s$^{-1}$] using the 2019 \xmm\ flux \citep{BJ2018}. Note that we varied $\dot{M}$ for the other \xmm\ and \nustar\ observation data with different X-ray fluxes. In addition, an uncertainty in fractional accretion column area ($f$), defined as $f = \dot{M}/4 \pi \dot{m} R^2$, leads us to consider a broad range of specific accretion rates for a robust mass measurement \citep{Vermette2023}. The maximum fractional accretion area, $f_{max} = \frac{R}{2 R_m}$, is found by assuming a dipole B-field geometry, where the infalling gas originates from the entire accretion disk \citep{2002apa..book.....F}. The effect of $\dot{m}$ on mass fitting depends primarily on its impact on $h$, and therefore, we select a range of accretion rates based on the resulting $h$ and corresponding value $f$. We present four different specific mass accretion rates of $\dot{m} = 0.03$, 0.05, 0.1 and 1 [g\,cm$^{-2}$\,s$^{-1}$], increasing the specific accretion rate for the \nustar\ fit appropriately. Our consideration of $\dot{m}$ yields shock heights between $\sim 20 \%$ and $ < 1 \%$ of the WD radius, representing a robust range of cases for mass measurement (Table: \ref{tab:mcvspec_fit}). The calculated range of fraction accretion area based on the $\dot{m}$ range is $0.0001 \leq f \leq 0.01$, encompassing the theoretical range of f ($0.001 \simlt f \simlt 0.02$) given by \cite{1988Rosen}. Furthermore, two IPs with measured or constrained $f$, XY Ari with $f \leq 2 \times 10^{-3}$ \citep{XYAri} and EX Hya with $f = 7.3_{-4.0}^{+29.3} \times 10 ^{-4} $ \citep{EXHya}, have accretion areas firmly on the lower end of Rosen's theoretical range, further validating our $f$-range. We considered an $f$ range substantially smaller than the theoretically maximum value of $f_{max} = 0.25$. Large values of $f$ correspond to lower specific accretion rates, leading to taller accretion columns.  The shock height, however, cannot be greater than the magnetosphere radius as it would violate the assumption that X-rays are emitted from the accretion column. Thus, for our estimates of $\dot{M}$, any $f > 0.02$ (corresponding to any $\dot{m} < 0.02$ g\,cm$^{-2}$\,s$^{-1}$), will always result in $h_s > R_m$. Hence, we consider a range of accretion rates that guarantees the $h < R_m$ while accounting for the systematic error in our mass measurement given by the uncertainty of $\dot{m}$ and $f$. 

The dual effects of varying $\dot{m}$ and $R_m/R$ substantially change the shock velocity of a WD. A lower specific accretion rate ($\dot{m}$) increases the accretion column height as suggested by \cite{Oliveira} for \src. 
A combination of the taller accretion column and smaller $R_m/R$ decreases $v_{ff}$ and $T_s$, and thus softens the X-ray spectrum because the free-falling distance ($R_m-h_s$) is reduced. In Figure \ref{fig:RmR} (left), we demonstrate how the shock temperature varies with a range of $\dot{m}$ and $R_m/R$, assuming the mean WD mass of 0.8 $M_\odot$ for CVs \citep{Pala2022}. Note that the shock temperature remains nearly constant when $\dot{m} \simgt 1$ g\,cm$^{-2}$\,s$^{-1}$ and this is consistent with a more detailed investigation presented by \citet{Hayashi2014}.

A tall accretion column minimizes X-ray reflection from the WD surface because the viewing angle of the WD from the shock height is significantly smaller. Thus, a tall shock height may account for the insignificant neutral 6.4 keV Fe line observed during the \xmm\ observations in our phenomenological analysis. The weakness of the 6.4 keV Fe line may instead be due to the viewing angle or low Fe abundance, however, its strong presence in the \nustar\ observation implies that the accretion column was shorter then due to greater mass accretion than in the other observations. Therefore, for completeness in our physically motivated model, we include {\tt reflect} to account for X-ray reflection off the WD surface. The shock height is calculated in each spectral fitting by the {\tt MCVSPEC} model. We incorporate the height into the X-ray reflection model ({\tt reflect}) via the reflection fraction factor defined by $\Omega/2\pi$ where $\Omega$ is the solid angle of the WD surface viewed from the shock height \citep{Tsujimoto2018}. We also ensured the condition of $R_m > h_s$ is satisfied in our spectral fitting to validate our assumption that X-rays are emitted from the accretion column.

Overall, our complete IP spectral model is given as {\tt tbabs*reflect*(MCVSPEC+gauss)}. As in the nmCV case, we fit the normalization of the Gaussian component of the \nustar\ spectra separately from the other observations. We also increase the specific accretion rate associated with the \nustar\ spectra by a factor of 4 according to the increase in unabsorbed flux. Otherwise, the model parameters are jointly fit. Table \ref{tab:mcvspec_fit} presents all our spectral fitting results in 4 cases. To reiterate, we considered four different specific mass accretion rates of $\dot{m} = 0.03$, 0.05, 0.1 and 1 g\,cm$^{-2}$\,s$^{-1}$, making up each of the four cases respectively. Case 2 and Case 3 represent the best-fit cases ($\chi_{\nu}^2 = 1.13$). At these accretion rates, the mass is constrained between 0.76 and 0.87 $M_{\odot}$. $R_m/R$ remained at $\sim 2$ in all cases however, for accretion rates past $0.1$ g\,cm$^{-2}$\,s$^{-1}$ the mass measurement remained at $\sim 0.84\; M_{\odot}$ as the shock height shrunk, but the quality of fit worsened ($\chi_{\nu}^2 = 1.32$ for case 4 and greater for higher accretion rates. Meanwhile, accretion rates below $0.05\;g\,cm^{-2}\,s^{-1}$, the mass decreased while the fit worsened ($\chi_{\nu}^2 = 1.35$ for case 1). Further decreasing the accretion rate resulted in taller accretion columns but also decreased the quality of fit and in some cases violated the assumption that the $h < R_m$. In our case, We constrained $R_m/R\approx 2$ and $M \approx 0.8 M_{\odot} $ as represented in Case 2-3 of Table \ref{tab:mcvspec_fit}. Considering the X-ray flux variability over the \xmm\ and \nustar\ observations, we derived the total mass accretion rate is in the range of $\dot{M} = (2.3\rm{-}9.0)\times10^{14}$ g\,s$^{-1}$ based on $L_X = G M \dot{M}\left(\frac{1}{R}-\frac{1}{R_m}\right)$ and $R_m/R = 2$ and $M = 0.8 \; M_{\odot}$. Finally, our fit results, along with the constrained $\dot{M}$ range from the known source distance, provide the means to bound the WD magnetic field strength. Following \citet{Vermette2023} and using the magnetic radius formula from \citet{Norton2004}, we determined $B \sim 0.1 \rm{-} 1$ MG.
\mycomment{
\begin{deluxetable*}{lcccccc}[ht]
\tablecaption{{\tt MCVSPEC} Fits to X-ray Spectra}
\label{tab:mcvspec_fit}
\tablecolumns{7}
\tablehead{
\colhead{Parameter}   
&
\colhead{Case 1}  
& 
\colhead{Case 2}
&
\colhead{Case 3}  
& 
\colhead{Case 4}
&
\colhead{Case 5}  
& 
\colhead{Case 6}
}
\startdata  
$N^{(i)}_H [10^{20} \rm{cm}^{-2}]^a\;*$ & 2.4 & 2.4 & 2.4 & 2.4 & 2.4 & 2.4  \\
$R_m/R$* & 2 & 2 & 2 & 8 & 8 & 8\\
$\dot{m}_{XMM}$ [g\,cm$^{-2}$\,$s^{-1}$]* & 0.05 & 0.1 & 1 & 0.05 & 0.1 & 1\\
$\dot{m}_{NU}$ [g\,cm$^{-2}$\,$s^{-1}$]* & 0.20 & 0.4 & 4 & 0.20 & 0.4 & 4\\
$M [M_{\odot}]$ & $0.84_{-0.05}^{+0.03}$ & $0.80_{-0.10}^{+0.04}$ & $0.72_{-0.05}^{+0.04}$ & $0.56_{-0.08}^{+0.04}$ & $0.53_{-0.08}^{+0.04}$ & $0.51_{-0.06}^{+0.03}$\\
$Z^b (Z_\odot)$ & $0.49_{-0.09}^{+0.10}$ & $0.50\pm0.09$ & $0.49_{-0.09}^{+0.10}$ & $0.48_{-0.12}^{+0.10}$ & $0.48_{-0.12}^{+0.10}$ & $0.47_{-0.11}^{+0.10}$\\
$EW^c_{\rm line}$ (eV) & $23_{-23}^{+50}$ & $28_{-28}^{+49}$ & $22_{-22}^{+50}$ & $26_{-26}^{+49}$ & $23_{-23}^{+49}$ & $19_{-19}^{+49}$  \\
$refl_{XMM}$* & 0.32 & 0.48 & 0.80 & 0.39 & 0.50 & 0.82 \\
$refl_{NU}$* & 0.56 & 0.67 & 0.90 & 0.61 & 0.71 & 0.90 \\
$cos(i)$ & $0.05_{-0.00}^{+0.25}$ & $0.05_{-0.00}^{+0.29}$ & $0.05_{-0.00}^{+0.17}$ & $0.05_{-0.00}^{+0.90}$ & $0.05_{-0.00}^{+0.90}$ & $0.05_{-0.00}^{+0.44}$ \\
$F_X [10^{-13}$ \fluxcgs]$^d$  & 6.87 & 6.75 & 6.84 & 6.73 & 6.78 & 6.85 \\
$F_X [10^{-13}$ \fluxcgs]$^e$  & 45.9 & 45.4 & 45.5 & 45.4 & 45.5 & 45.5 \\
$\chi^2_{\nu}$ (dof) & 1.15 (385) & 1.16 (385) & 1.15 (385) & 1.15 (385) & 1.15 (385) & 1.15 (385) \\
$h_s/R_{XMM}$ & 37 \% & 17 \% & 2 \% & 26 \% & 15 \% & 1 \% \\
$h_s/R_{NU}$ & 12 \% & 6 \% & 1 \% & 8 \% & 5 \% & 1 \% \\
\enddata
All errors shown are $90 \%$ confidence intervals. The Gaussian line was not significantly detected in any fit. \\
$^a$ The ISM hydrogen column density is associated with {\tt tbabs}, which is multiplied to all the models. \\
$^b$ Abundance relative to solar. \\
$^c$ The equivalent width of the Gaussian component with $E =  6.4$ keV and $\sigma = 0.01$ keV. \\
$^d$ 3 -- 10 keV flux of the \XMM\ data. \\
$^e$ 3 -- 10 keV flux of the \nustar\ data. \\
$^*$ The parameter is frozen. \\

\end{deluxetable*}
}
\begin{deluxetable*}{lcccccc}[ht]
\tablecaption{{\tt MCVSPEC} Fits to X-ray Spectra}
\label{tab:mcvspec_fit}
\tablecolumns{5}
\tablehead{
\colhead{Parameter}   
&
\colhead{Case 1}  
& 
\colhead{Case 2}
&
\colhead{Case 3}  
&
\colhead{Case 4}  
}
\startdata  
$N^{(i)}_H [10^{20} \rm{cm}^{-2}]^a\;*$ & 4.2 & 4.2 & 4.2 & 4.2  \\
$\dot{m}_{XMM}$ [g\,cm$^{-2}$\,$s^{-1}$]* & 0.03 & 0.05 & 0.1 & 1 \\
$\dot{m}_{NU}$ [g\,cm$^{-2}$\,$s^{-1}$]* & 0.12 & 0.20 & 0.4 & 4 &\\
$M [M_{\odot}]$ & $0.69\pm0.01$   & $0.77\pm0.01$ & $0.84\pm0.03$ & $0.83_{-0.02}^{+0.03}$ \\
$Z^b (Z_\odot)$ & $0.20_{-0.06}^{+0.07}$  & $0.30\pm0.08$ & $0.42\pm0.11$& $0.51_{-0.13}^{+0.15}$ \\
$EW^c_{\rm line}$ (eV) & $38_{-38}^{+49}$  & $10_{-10}^{+45}$ & $0_{-0}^{+36}$ & $0\pm0$  \\
$EW^d_{\rm line}$ (eV) & $270\pm 77$  & $229\pm73$ & $241\pm74$ & $342\pm82$ \\
$refl_{XMM}$* & 0.45  & 0.42 & 0.46 & 0.76 \\
$refl_{NU}$* & 0.64  & 0.62 & 0.64 & 0.87 \\
$cos(i)$ & $0.995_{-0.823}^{+0.000}$ & $0.995_{-0.822}^{+0.000}$ &  $0.995_{-0.995}^{+0.000}$ & $0.995_{-0.995}^{+0.000}$ \\ 
$F_X [10^{-13}$ \fluxcgs]$^e$ & 6.48 & 6.79 & 7.06 & 7.68 \\
$F_X [10^{-13}$ \fluxcgs]$^f$ & 39.1 & 40.0 & 38.9 & 35.0 \\
$\chi^2_{\nu}$ (dof) & 1.35 (630)  & 1.13 (630) & 1.13 (630) & 1.32 (630) \\
$R_m/R$ & $1.61_{-0.03}^{+0.02}$  & $1.82 \pm 0.03$ & $2.02_{-0.03}^{+0.10}$ & $1.99_{-0.06}^{+0.10}$ \\
$h_s/R_{XMM}$ & 20\%   & 22\% & 19\% & 3\% \\
$h_s/R_{NU}$ & 7\%   & 8\% & 7\% & 1\% \\
\enddata
All errors shown are $90 \%$ confidence intervals. The Gaussian line was not significantly detected in any fit. \\
$^a$ The ISM hydrogen column density is associated with {\tt tbabs}, which is multiplied to all the models. \\
$^b$ Abundance relative to solar. \\
$^c$ The equivalent width of the Gaussian component with $E =  6.4$ keV and $\sigma = 0.01$ keV for the both \xmm\ datasets. \\
$^d$ The equivalent width of the Gaussian component with $E =  6.4$ keV and $\sigma = 0.01$ keV for the \nustar\ data. \\
$^e$ 3 -- 10 keV flux of the \xmm\ 2019 data. \\
$^f$ 3 -- 10 keV flux of the \nustar\ data. \\
$^*$ The parameter is frozen. \\
\end{deluxetable*}
\begin{figure}[ht]
    \centering
    \begin{minipage}{0.48\linewidth}
        \includegraphics[width=\linewidth]{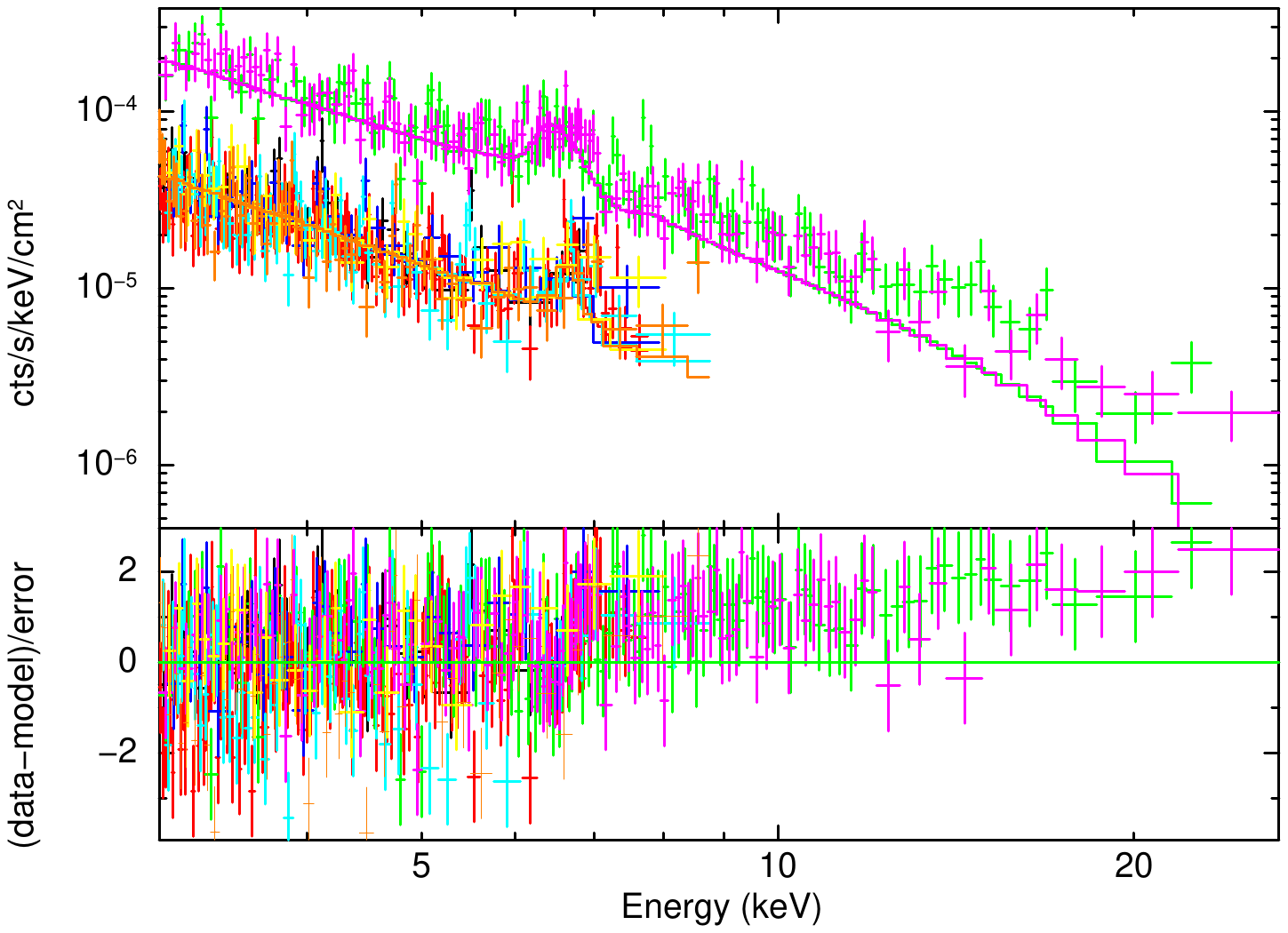}
    \end{minipage}
    \hfill
    \begin{minipage}{0.48\linewidth}
        \includegraphics[width=\linewidth]{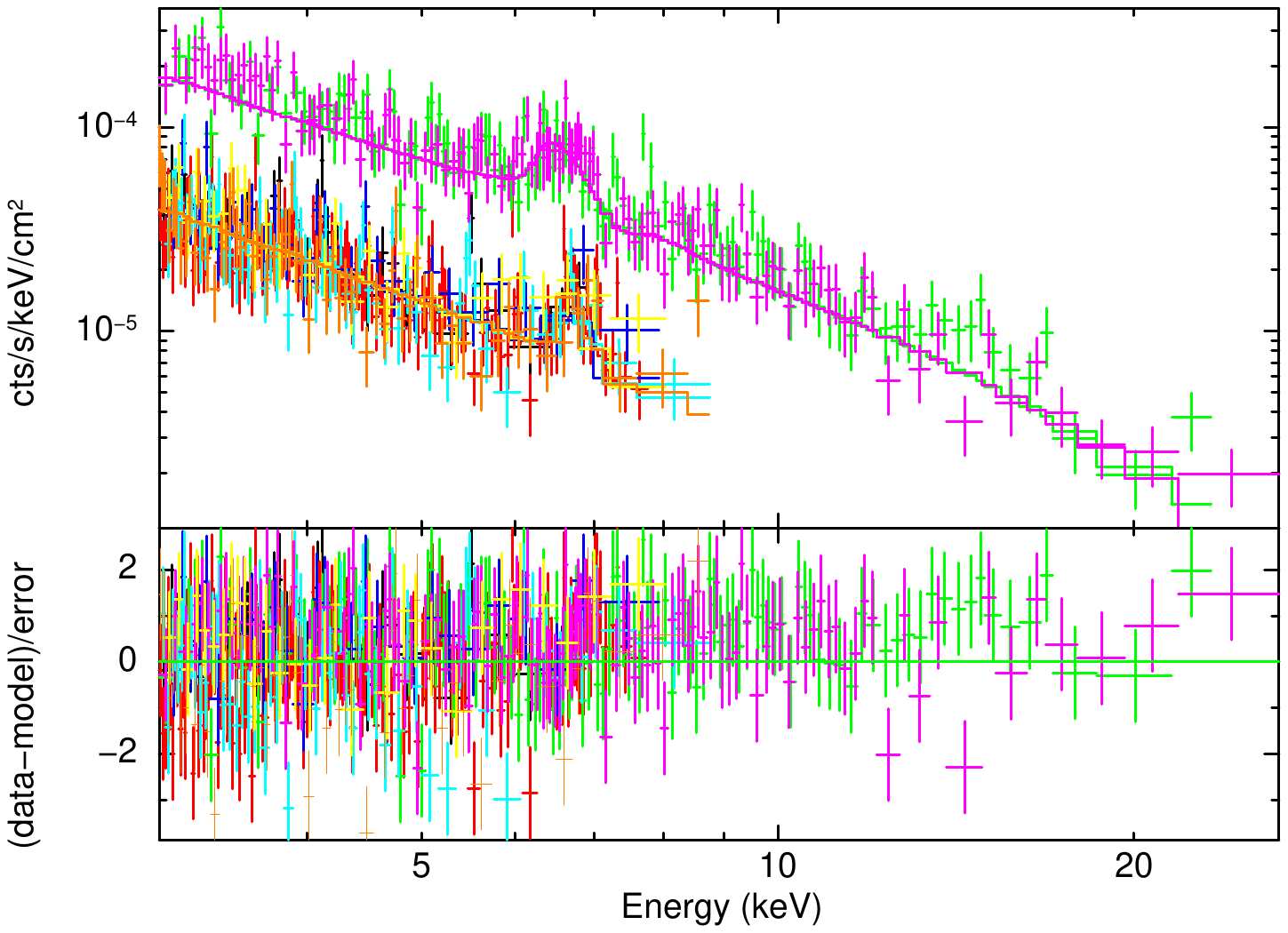}
    \end{minipage}

    \begin{minipage}{0.48\linewidth}
        \includegraphics[width=\linewidth]{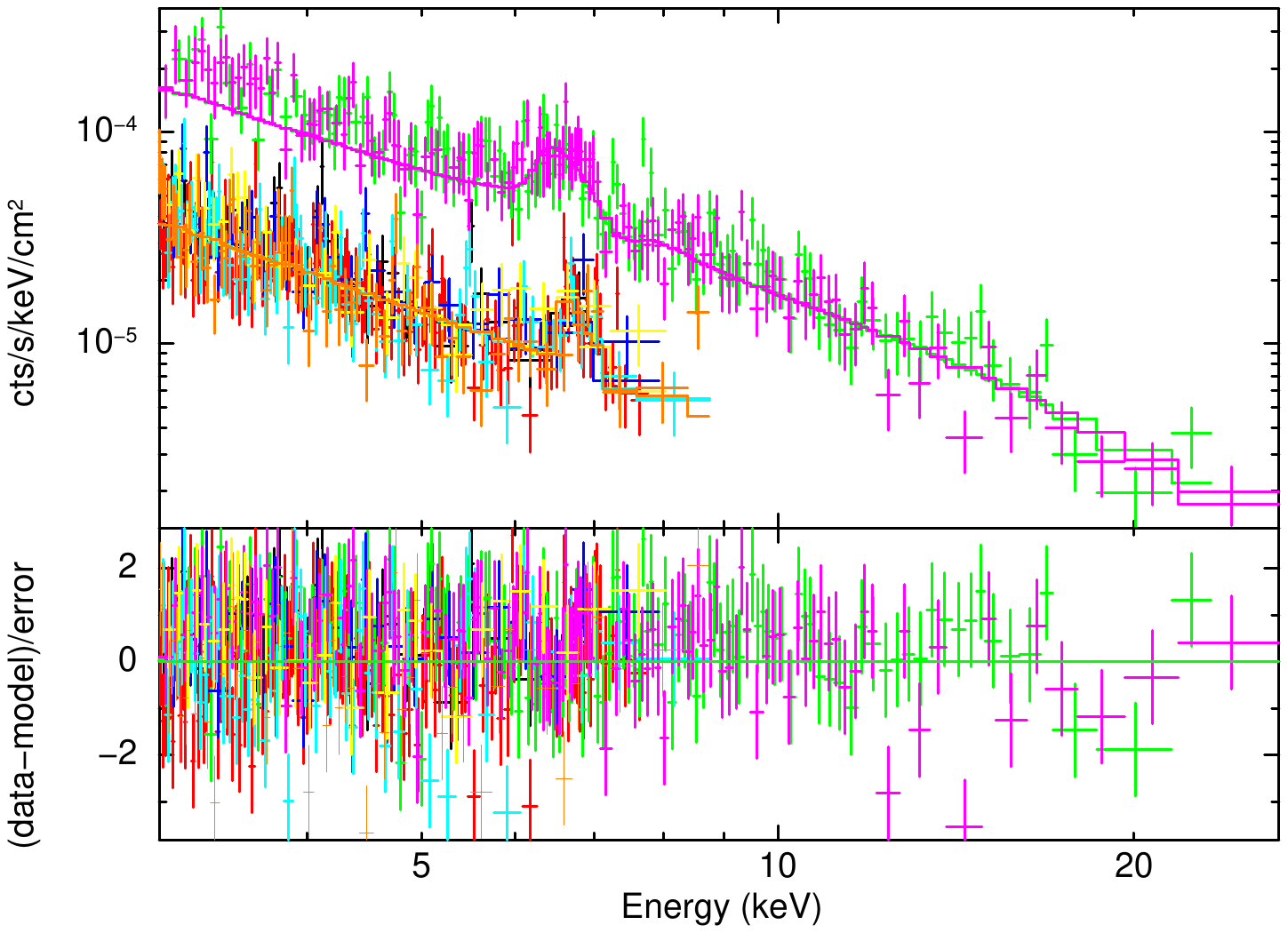}
    \end{minipage}
    \hfill
    \begin{minipage}{0.48\linewidth}
        \includegraphics[width=\linewidth]{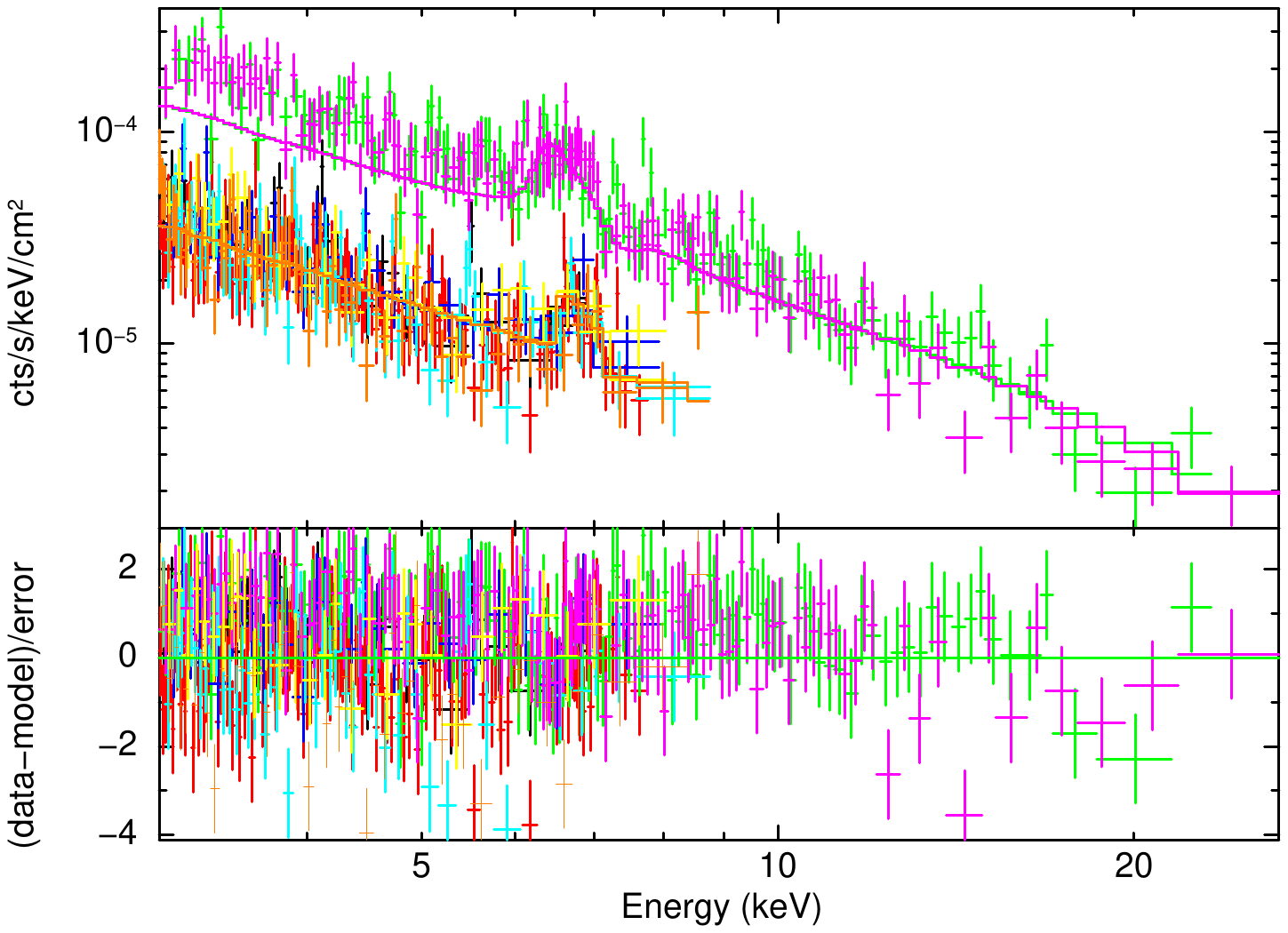}
    \end{minipage}

    \caption{\xmm\ and \nustar\ spectra and residuals of \src\ fit with the {\tt tbabs}*({\tt mcvspec}+{\tt gauss}) model. The cases are presented as follows: case 1 (upper left), case 2 (upper right), case 3 (lower left), case 4 (lower right). The 2019 \xmm\ observation (obsID = 0842570101) is presented in the 0.3 - 10 keV band with MOS1 in blue, MOS2 in yellow, and pn in black. Similarly, 2022 \xmm\ observation (obsID = 0902500101) is presented in the 0.3 - 10 keV band with MOS1 in cyan, MOS2 in orange, and pn in red. The \nustar\ (obsID = 30801006002) observation is presented in the 3-30 keV band, with FPMA in green and FPMB in magenta.}
    \label{fig:mcvspec_fits}
\end{figure}
\section{Discussion} \label{sec:disc}
We have comprehensively investigated the fast-spinning CV, \src, through follow-up \xmm, \nice, and \nustar\ observations, providing deeper insight into the X-ray properties of \src. Below, we discuss our X-ray analysis results and their implications for \src\ and, more broadly, FSCVs and LLIPs. 

\subsection{X-ray timing properties}

The 29.6 second period was stable in the X-ray band across all \xmm\ and \nice\ observations, consistent with the properties of a WD spin period. Although not measured, the period derivative was constrained to $|\dot{P}| < 1.8\times$ $10^{-12}$ s/s.  Our constraints of $\dot{P}$ confirm that \src\ has a smaller spin derivative than WZ Sge, but it is still above the range of the other FSCVs, which demonstrate period derivatives of $\sim 10^{-13}\rm{-}10^{-14}$ s/s (Table: \ref{tab:src_types}). V1460 Her's spin derivative has not been measured with a stringent upper limit of $|\dot{P}| < 3 \times 10^{-14}$ s/s, smaller than any of the other FSCVs \citep{PelisoliIP}. Its spin-down timescale of $4 \times 10^7$ yr is much longer than AE Aqr's and AR Sco's timescales, $10^7$ yr \citep{Dejager1994} and $5\times 10^6$ \citep{2018AJ....156..150S}, respectively; thus, it was theorized to be in a state of quasi-equilibrium rather than spinning up or spinning down \citep{PelisoliIP}. Interestingly, in contrast to many IPs, other FSCVs (e.g.,  AE Aqr, WZ Sge) have been spinning down. The spin evolution of FSCVs may be related to why all known FSCVs are LLIPs. FSCVs must have undergone substantial mass accretion in the past to be spun up to their current periods and would represent a late stage of CV evolution \citep{Zorotovic2011}. Their donor may have less material to feed to the WD, leading to decreased accretion rates and luminosity. Future observations with a more sensitive $\dot{P}$ measurement will be essential in determining whether \src\ is currently spinning up or down. 

The hardness ratios presented in the timing analysis and the spectral variability demonstrated in the later phase-resolved spectroscopy imply the presence of an accretion curtain, a fundamental feature of IPs. In IPs, the softening of emission at the peak of the pulse profile of its spin period is due to the accretion curtain \citep{1988Rosen}. Accreting matter flows along the magnetic field lines of the WD, forming an arc of material, the curtain, that leads into the column. Due to the WD rotation, the accretion curtain may absorb soft X-ray photons and obscure the column from the observer, causing spin modulation. \cite{Oliveira} excluded a complex absorber to model the X-ray spectra of \src, noting that the lack of such a feature is common to LLIPs and provides further evidence of a tall accretion column.

Firstly, the hardness ratio with our highest quality data set demonstrates modulation at the spin period with 3$\sigma$ significance. As per the accretion curtain model, the X-ray emission is obscured when the curtain passes between the observer and the column. Necessarily, this hardens the spectra as the curtain absorbs soft X-ray emission most. The same effect is visible in the change in temperature in the phase-resolved spectral analysis, where the fourth phase bin in Figure \ref{fig:phasecuts} has the fewest source counts yet the highest temperature. Naturally, the increase in soft X-ray flux corresponds with the increase in counts, finding a maximum in the second phase bin. These results support \src's classification as an IP, indicating an accretion column that could not be previously confirmed. Despite this, the relationship between $n_H$ and soft X-ray flux remains inconclusive. Typically, IPs demonstrate an anti-correlating phase dependence between soft X-ray flux and absorption. This is because X-ray absorption should be more significant at spin phases where the curtain blocks soft X-ray photons, as reported by \citet{Joshi2019}. However, no such anti-correlation was clearly detected in our phase-resolved spectral analysis. 

\subsection{X-ray spectral properties} 

Our broadband X-ray spectral analysis demonstrated that the spectrum of \src\ is best described by a multi-temperature thermal emission model with three {\tt APEC} components with $kT \sim$ 0.3, 1, and 5 keV in the "low" flux state observed by \xmm\ and \nice. Meanwhile, the "high" state, observed in the 3 - 30 keV band, fits best to a single-temperature thermal emission model with $kT = 8.4$ keV.  These spectral properties are similar to those of other FSCVs previously studied in the X-ray band with ample photon statistics, such as AE Aqr and WZ Sge. A broadband spectral analysis of AE Aqr's X-ray emission is well-fit by both a three-temperature thermal plasma ($kT = 0.75, 2.3$ and 9.3 keV) or two thermal plasma components and a power-law component ($kT = 1.0, 4.6$ keV; $\Gamma$ = 2.5) \citep{Kitaguchi_2014}. The spectrum of WZ Sge also fits well to a multi-temperature plasma model with $kT = 1.3$ and $9.0$ keV. Unlike \src, a 3rd thermal plasma component was not required in the most updated study of its X-ray emission \citep{Nucita2014}. 

Unlike most IPs, the Fe K-$\alpha$ emission line at 6.4 keV is not detected when we fit the 3-temperature plasma model with a Gaussian component (Table \ref{tab:wonu_prelimfits}). If \src\ is an IP with a truncated accretion disk, the lack of the 6.4 keV Fe line indicates a tall accretion column in which case hard X-rays emitted from the shock region are less likely to be reflected off the WD surface \citep{Tsujimoto2018}. In the high state, the Fe line is substantially stronger with EW = $174_{-76}^{+75}$ eV. An enhanced accretion rate in an IP would increase the total amount of X-ray emission while simultaneously shortening the accretion column causing more X-rays to be reflected off the WD surface and strengthening the Fe line. Our best {\tt mcvspec} fits further demonstrate a substantial difference in accretion column height with $\frac{h}{R} \sim 20 \%$ and $\frac{h}{R} \sim 8 \%$ in the low flux and high flux states, respectively. Should the source be an nmCV, our weak iron line is consistent with other dwarf novae such as V893 Sco, whose 6.4 keV Fe line is weak with EW = $45_{-12}^{+11}$ eV \citep{Byckling2010}. Furthermore, should the high state represent a dwarf novae outburst, the increased strength of the Fe line is similar to how the Fe line becomes prominent in the X-ray spectra of WZ Sge during outburst \citep{balman2014inner}. If \src\ is undergoing an outburst, however, it joins two other DNe, GW Lib and UW Gem, in demonstrating an increase in X-ray luminosity \citep{BycklingGW,guver}. Many other DNe, including WZ Sge, show decreased X-ray flux while optical flux increases during outburst \citep{balman2014inner}. In summary, the weak 6.4 keV Fe line seems to be common among the FSCVs, including \src. 

In addition, FSCVs demonstrate softer X-ray spectra ($kT \simlt$ 10 keV) compared to typical IPs ($kT \sim 20{\rm-}40$ keV). In both our high and low states, our phenomenological fits indicated a low peak temperature. Traditionally, CV and mCV spectral models, such as {\tt MKCFLOW} and {\tt ipolar}, measure WD mass by correlating it with shock temperature \citep{Yu2018, Suleimanov2016}. As mentioned above, smaller $R_m/R$ (due to faster WD spins) and smaller $\dot{m}$ (due to fainter X-ray emission) will result in lower $T_s$ as demonstrated by Figure \ref{fig:RmR} (left). If a fast-spinning object is in spin equilibrium, where $R_m/R \sim P^{2/3}$, the magnetosphere radius will be low compared to typical IPs, whose spin periods in the hundreds of seconds result in $R_m/R = 8\rm{-}25$ \citep{Suleimanov2016}. Therefore, if the FSCVs and LLIPs possess truncated accretion disks, they should exhibit softer X-ray spectra than other more slowly spinning IPs.

\subsection{WD mass and B-field measurements} 

Although hampered by the unknown source type, our WD mass measurement of \src\ yielded $M = 0.7\rm{-}1.0\, M_\odot$, more specifically $M = 0.84\rm{-}1.03\, M_\odot$ and $M = 0.76\rm{-}0.87\, M_\odot$ for the nmCV and IP cases, respectively. With the benefit of broadband X-ray data (0.3 - 30 keV instead of 0.3 - 10 keV), improved photon statistics from four additional X-ray observations, and the use of physically motivated models, our results represent the first accurate WD mass determination of an FSCV using X-ray data. While previous mass estimates demonstrated a substantially lower mass \citep{Oliveira}, our WD mass range is above the theoretical centrifugal break-up threshold of $0.56 M_\odot$ \citep{Otoniel_2021}. Should \src\ be an IP, our results indicate $B = 0.1 \rm{-} 1$ MG. Should the source be an nmCV, however, the WD B-field should be lower. 
As mentioned by \cite{Oliveira}, \src\ was already predicted to have a weak B-field as necessitated by its low magnetospheric radius and accretion rate.

\src\ is the second known FSCV below the period gap \citep{Rodrigues2023}. As previous studies have suggested, mass accretion should have increased both the angular momentum and the WD mass substantially, especially for evolved CVs below the period gap \citep{Zorotovic2011}. \src\ must have undergone substantial spin-up in the past to reach its current fast period. CVs above the period gap are expected to be more massive than those below it, as they must be sufficiently massive pre-CVs to initiate the stable mass transfer. Yet, the WD mass of \src\ is close to the mean range of CVs \citep{Pala2022}. It is still possible that episodic nova eruptions may have kept the WD mass of \src\ relatively smaller than other massive IPs (e.g, \citet{Gerber2024}). It is thus important to measure WD masses from other FSCVs and find out whether any correlation is present between $M$, spin, and orbital periods. 

\section{Conclusion} \label{sec:conc}

We present the most thorough X-ray analysis of \src\ with broadband \XMM, \nustar, and \nice\ observations collected over three years. Our results demonstrate that \src\ is an important source for probing the properties of a rare class of FSCVs in the X-ray band. As with other FSCVs, a soft multi-temperature thermal plasma model best describes its broadband X-ray spectra. We confirm the lack of a Fe-K-$\alpha$ line in our phenomenological and nmCV models in the low state, while detecting one during the outburst observed by \nustar\ . \src's IP-like behavior, including its accretion-powered emission and evidence of an accretion column, combined with its nmCV properties and weak Fe line in its "low" state make its classification uncertain. Nevertheless, we constrained \src's fundamental properties, presenting the first reliable WD mass measurement of a FSCV using X-ray spectral data, yielding $M = 0.7 \rm{-} 1.0\, M_{\odot}$ within the typical WD mass range of CVs. \src's IP-like behaviors are in contrast to the other FSCVs' unique properties and raise more questions about their evolutionary paths and emission mechanisms (Table \ref{tab:src_types}). Our companion paper (van Dyk et al., in preparation) will discuss the optical properties of \src\, which are similar to those of WZ Sge-type stars with dwarf novae, superoutbursts, and superhumps.

LLIPs may not be rare objects, and there may be a substantial population of LLIPs that will be discovered as more sensitive X-ray surveys become capable of probing low-luminosity sources \citep{Pretorius}. The ongoing {\it eROSITA} survey may uncover more LLIPs in the X-ray band.  Meanwhile, the {\it ZTF}, and {\it Rubin} optical surveys of known CVs may identify more FSCVs and discover new sources of interest \citep{2019cwdb.confE..49S}. In the near future, the {\it HEX-P} X-ray probe mission, which covers a broad X-ray band of 0.1--80 keV, will provide the sensitivity to obtain a high-quality X-ray spectrum of these sources \citep{Mori2024}. A larger sample size of FSCVs and LLIPs will illuminate the origin of unique sources like AR Sco and AE Aqr and answer whether this emergent class of objects is populated by typical mCVs or rare systems like WD pulsars or CV propellers. 
    
\section{Acknowledgments}
Support for this work was provided by NASA through \nice\ Cycle 3 (NNH20ZDA001N-NICER), \xmm\ Cycle 21 (XMMNC21) Guest Observer program, and NASA ADAP program (NNH22ZDA001N-ADAP) grants. RLO is a Research Fellow of the Brazilian institution CNPq (PQ-315632/2023-2).

\software{{\tt astropy} \citep{astropy}, {\tt stingray} \citep{matteo_bachetti_2022_6394742},
HEASoft Version 6.25 \citep{heasoft}, FTools Version 6.25 \citep{heasoft}}

\bibliography{main}{}

\end{document}